\documentclass[]{elsarticle}

\usepackage{hyperref}
\usepackage{graphicx}

\newtheorem{remark}{Remark}[section]
\newtheorem{theorem}{Theorem}[section]

\usepackage{amsmath,amssymb,bm,url,color}

\newcommand{\plusR}{\mathbb{R}^+}

\newcommand{\diff}{{\rm\,d}}
\newcommand{\equaref}[1]{(\ref{eq:#1})}

\newcommand{\myfig}[3]{
  \begin{figure}[thb]
  \centering
  \includegraphics[width=#1.0cm, clip=true]{#2.pdf}
  \caption{#3\label{fig:#2}}
  \end{figure}
}

\journal{Annual Reviews in Control}

\begin{document}

\begin{frontmatter}

\title{A time-modulated Hawkes process to model the spread of COVID-19 and the impact of countermeasures\tnoteref{mytitlenote}}
\tnotetext[mytitlenote]{The simulation code and data used in this work are available under GPL v3 on GitHub:
\href{https://github.com/michelegaretto/covid.git}{\url{https://github.com/michelegaretto/covid.git}}}


\author[unito]{Michele Garetto\corref{mycorrespondingauthor}}
\cortext[mycorrespondingauthor]{Corresponding author}
\ead{michele.garetto@unito.it}

\author[polito]{Emilio Leonardi}
\ead{emilio.leonardi@polito.it}

\author[cnr]{Giovanni Luca Torrisi}
\ead{giovanniluca.torrisi@cnr.it}

\address[unito]{Universit\`{a} degli Studi di Torino, C.so Svizzera 185, Torino, Italy}
\address[polito]{Politecnico di Torino, C.so Duca degli Abruzzi 24, Torino, Italy}
\address[cnr]{IAC-CNR, Via dei Taurini 19, Roma, Italy}

\begin{abstract}
Motivated by the recent outbreak of coronavirus (COVID-19), we propose
a stochastic model of epidemic temporal growth and mitigation
based on a time-modulated Hawkes process. The model is sufficiently rich to
incorporate specific characteristics of the novel coronavirus,
to capture the impact of undetected, asymptomatic and super-diffusive individuals, and
especially to take into account time-varying counter-measures and detection efforts.
Yet, it is simple enough to allow scalable and efficient computation of
the temporal evolution of the epidemic, and exploration of what-if scenarios.
Compared to traditional compartmental models, our approach 
allows a more faithful description of virus specific features, such as
distributions for the time spent in stages, which is crucial 
when the time-scale of control (e.g., mobility restrictions) is comparable 
to the lifetime of a single infection. We apply the model to the 
first and second wave of COVID-19 in Italy,
shedding light onto several effects related to mobility restrictions
introduced by the government, and to the effectiveness of contact tracing
and mass testing performed by the national health service.     
\end{abstract}




\end{frontmatter}


\section{Introduction and related work}
Models of epidemic propagation are especially useful when they provide key insights 
while retaining simplicity and generality.
For example, in mathematical epidemiology the SIR model 
reveals in very simple terms the fundamental role of the basic reproduction number ($R_0$)
which governs the macroscopic, long-term evolution of the outbreak in a homogeneous population.
The vast majority of models developed for the novel SARS-CoV-2
are extensions to the classic SIR, along the standard approach
of introducing additional compartments to describe different phases 
of the infection, the presence of asymptomatic, symptomatic or pauci-symptomatic individuals, 
the set of quarantined, hospitalized people, and so on. Such models lead to
a system of coupled ODE's with fixed or time-varying coefficients to be estimated from traces.
A very incomplete list of modeling efforts pursued in this direction,  
early applied to COVID-19, includes the SEIR models in \cite{Fang2020,KUCHARSKI2020553,columbia,ucla,Miller},
the SIRD models in \cite{Anastassopoulou,CALAFIORE}, the SEPIA model in \cite{gatto},
the SIDHARTE model in \cite{Giordano2020}.  
     
In this paper we develop a slightly more sophisticated model that allows a more accurate
representation of native characteristics of a specific virus, such as actual distributions
of the duration of incubation, pre-symptomatic and symptomatic phases, for various
categories of infected.
 This level of detail is important when the intensity of applied 
countermeasures varies significantly over time-scales comparable to that 
of an individual infection, and we believe it is essential to address fundamental questions such as: i) when and to what extent can we expect to see the effect of specific mobility restrictions introduced by a national government at a given point in time? ii) what is the impact of hard vs partial lockdowns enforced for given numbers of days? when can restrictions be safely released to restart economic and social activities while still keeping the epidemic under control? 

Specifically, we propose {\color{black} and analyse} a novel, modulated version of the {\color{black} marked} Hawkes process, a self-exciting 
stochastic point process with roots in geophysics and finance \cite{hawkes}.
In a nutshell, in the standard {\color{black} marked Hawkes process each event $i$ with mark $m_i$, occurring at time $t_i$,
generates new events with stochastic intensity $\nu(t-t_i, m_i)$, where $\nu( \cdot, m_i)$ is a generic
kernel. The process unfolds through successive generations of events, starting
from so-called immigrants (generation zero). }   
In our model events represent individual infections, and a modulating function $\mu(t)$,
which scales the overall intensity of the process at real time $t$, allows us to take into account
the impact of time-varying mobility restrictions and social distance limitations.
In addition, we model the transition of infected, undetected individuals to a quarantined state at 
inhomogeneous rate $\rho(t)$, to describe the time-varying effectiveness of contact tracing and
mass testing.    

We mention that branching processes of various kind, including Hawkes processes,  
have been proposed in various biological contexts \cite{harris,mei,Kim401,xu17}.
In \cite{akian2020probabilistic}, a probabilistic extension of (deterministic) 
discrete-time SEIR  models,  based on multi-type branching processes, has been recently
applied to COVID-19 to capture the impact of detailed distributions of the time spent in different phases,
together with mobility restrictions and contact tracing.
In parallel to us, authors of \cite{chiang} have proposed a Hawkes process
with spatio-temporal \lq\lq covariates" for modeling COVID-19 in the US, together with 
an EM algorithm for parameter inference. The work in \cite{chiang} is somehow orthogonal to us,
since they focus on spatial and demographic features, aiming at predicting the
trend of confirmed cases and deaths in each county. In contrast, we introduce marks and stages
to natively model the course of an infection for different categories of individuals, 
with the fundamental distinction between real and detected cases. Moreover,
we take into account the impact of time-varying detection efforts and contact tracing.
At last, we obtain analytical expressions for the first two moments of the number of 
individuals who have been infected within a given time.

We demonstrate the applicability of our model to the novel COVID-19 pandemic
by considering real traces related to the first and second wave of coronavirus in Italy.
Our fitting exercise, though largely preliminary and based on incomplete information,
suggests that our approach has good potential and can be effectively used both for 
planning counter-measures and to provide an a-posteriori explanation of observed 
epidemiological curves.

The paper is organized as follows. In Section \ref{sec:model} we provide the mathematical
formulation of the proposed modulated Hawkes process to describe the temporal evolution of the epidemic.
In Section \ref{sec:compsir} we motivate our approach by comparing it to the standard SIR model.
Some mathematical results related to the moment generating function of our process
are presented in Section \ref{sec:moments}.
In Section \ref{sec:covidmodel} we describe our COVID-19 model based on the proposed 
approach. We separately fit our model to the first and second wave of COVID-19 in Italy
in Sections \ref{sec:firstwave} and \ref{sec:secondwave}, respectively, offering hopefully 
interesting insights about what happened in this country (and similarly in other European countries) during the recent pandemic.
We conclude in Section \ref{sec:conclusions}.

\section{Mathematical formulation of modulated Hawkes process}\label{sec:model}
We first briefly recall the classic Hawkes process restricted
to the temporal dimension (the spatio-temporal formulation is similar, but we focus in this work
on the purely temporal version). Events of the process (or points) 
occur at times $T=\{T_i\}_{i\geq 1}$, which are $\mathbb{R}$-valued random variables. 
A subset $I$ of these points, called immigrants, are produced by an inhomogeneous
Poisson process of given intensity $\sigma(t)$. Each immigrant, independently of others,
is the originator of a progeny (or cluster) of other points, dispersed in the future through a self-similar
branching structure: a first generation of points is produced with intensity \mbox{$\nu(t-T_j)$}, where
$T_j$ is the occurrence time of immigrant $j$, and \mbox{$\nu: \plusR \rightarrow \plusR$} is a kernel function.
Each point of the first generation, in turn, generates new offsprings in a similar fashion, 
creating the second generation of points, and so on.   

The above process can be easily extended to account for different types of points
with type-specific kernel functions. Types are denoted by marks $M=\{M_i\}_{i\geq 1}$, which are
assumed to be i.i.d. random variables with values on an arbitrary measurable space $(\mathrm{M},\mathcal{M})$,
with a probability distribution $\mathbb{Q}$. Let $N(t,m)$ be the counting process associated  
to the marked points $N=\{(T_i,M_i)\}_{i\geq 1}$.
The (conditional) stochastic intensity $\lambda(t)$ of the overall process is then given by:
\begin{eqnarray*}\label{eq:haw}
\lambda(t)  =   \sigma(t) + \int_{0}^t \nu(t-s,m)  N(\diff s, \diff m)  = \sigma(t) + \sum_{{T_k} \cap (0,t) } \nu(t-T_k,M_k) 
\end{eqnarray*}
where $\nu: \plusR \times \mathrm{M} \rightarrow \plusR$ is a type-dependent kernel function.
In the following we assume $\overline{\nu}(t):=\mathbb{E}[\nu(t,M_1)]$ to be summable,. i.e.
\begin{equation}\label{hp:integrability}
R_0 = \int_0^\infty\overline{\nu}(t)\mathrm{d}t<\infty 
\end{equation}
Note that $R_0$ is the average number of offsprings generated by each point, which is
usually referred to as basic reproduction number in epidemiology. The process
is called subcritical if $R_0 < 1$, supercritical if $R_0 > 1$.

Our main modification to the above classic marked Hawkes process $N$
is to modulate the instantaneous generation rate of offsprings $N'=N\setminus I$ by a positive, bounded
function $\mu(t): \plusR \rightarrow \plusR$, representing the impact of mobility restriction countermeasures.
By so doing, we obtain the modified stochastic intensity of the process:
\begin{align}\label{eq:modhaw}
\lambda(t)  &=   \sigma(t) + \mu(t)\left[\int_{0}^t \nu(t-s,m)  N(\diff s, \diff m) \right] \nonumber \\
&=  \sigma(t)+\mu(t)\left[\sum_{{T_k} \cap (0,t) } \nu(t-T_k,M_k) \right] 
\end{align}
Note that, when $\mu(t) = \mu$ is constant, we re-obtain a classic Hawkes process
with modified kernel $\mu\,\nu()$. In general, the obtained process is no longer self-similar. 
In particular, the average number of offsprings generated by a point becomes a function
of time:
\begin{equation}\label{eq:Rt}
R(t) = \int_0^\infty \mu(t+\tau) \overline{\nu}(\tau) \diff \tau 
\end{equation}      
which provides the infamous real-time reproduction number usually referred to on the media as \lq Rt index'.  

We emphasize that the process can initially start in the supercritical regime, and then
it can become subcritical for effect of a decreasing function $\mu()$, a case of
special interest in our application to waves of COVID-19.
We mention that the great bulk of literature related to the Hawkes
process and its applications to geophysics and finance focuses on the subcritical regime, whereas our
formulation applies also to the supercritical regime, which is more germane to epidemics.  




The conditional intensity \equaref{modhaw} can be easily de-conditioned 
with respect to $N$, obtaining the \lq average' stochastic intensity
$\overline{\lambda}(t):=\mathbb{E}[\lambda(t)]$: 
\begin{align}\label{eq:barlambda}
\overline{\lambda}(t)&=\sigma(t)+\mu(t)\mathbb{E}\left[\int_{(0,t)\times\mathrm{M}}\nu(t-s,m)N(\mathrm{d}s,\mathrm{d}m)\right]\nonumber\\
&=\sigma(t)+\mu(t)\mathbb{E}\left[\int_{(0,t)\times\mathrm{M}}\nu(t-s,m)\lambda(s)\,\mathrm{d}s\mathbb{Q}(\mathrm{d}m)\right]\nonumber\\
&=\sigma(t)+\mu(t)\mathbb{E}\left[\int_{0}^{t}\overline{\nu}(t-s)\lambda(s)\mathrm{d}s\right]\nonumber\\
&=\sigma(t)+\mu(t)\int_{0}^{t}\overline{\nu}(t-s)\overline{\lambda}(s)\mathrm{d}s.
\end{align}
where we recall that  $\overline{\nu}(t)= \mathbb{E}[\nu(t,M_1)]$.
 
We observe that \equaref{barlambda} is a linear Volterra equation of the second kind,
which can be efficiently solved numerically.\footnote{For example, the standard trapezoidal rule
allows obtaining a discretized version of $\overline{\lambda}(t)$ by a matrix inversion.}   
In the special case of constant modulating function, \equaref{barlambda}
reduces to a convolution equation which can be analyzed and solved by means of
Laplace transform techniques (see \ref{app:explicit}).  

At last we introduce  the total number of points up to time $t$, $N(t)$ (regardless of their associated marks), and its average:
\begin{equation}\label{eq:barN}
\overline{N}(t) = \int_0^t  \overline{\lambda}(\tau) \diff \tau
\end{equation}

\section{Comparison with SIR model}\label{sec:compsir}
When the kernel function has an exponential shape, i.e., $\nu(t) = K e^{-\gamma t}$, it is possible to establish a simple connection between the Hawkes process and the classic SIR model \cite{australiani}.
Specifically, consider a stochastic SIR model with an infinite population of susceptible individuals, where each infected generates new infections at rate $\beta$, and recovers at rate $\gamma$ (i.e., after an exponentially distributed amount of time of mean $1/\gamma$). 
Then the average intensity of the process generated by this stochastic SIR, averaging out the times at which nodes recover,
has exactly the (conditional) intensity of a Hawkes process starting with the same number of initially
infected nodes, no further immigrants, and $\nu(t) = \beta e^{-\gamma t}$ \cite{australiani}.
This can be intuitively understood by considering that in SIR the average, effective rate with which an infected individual generates new infections after time $t$ since it became infected equals $\beta$ times the probability that the node has not yet recovered, which is  $e^{-\gamma t}$.

We remark that in \cite{australiani} authors push this equivalence a bit further, by showing that
a SIR model with {\em finite} population $N_0$ is equivalent to a modulated Hawkes process
similar to \equaref{modhaw}, where $\mu(t) = 1-N(t)/N_0$. In our model and its application
to COVID-19, we do not consider the impact of finite population size, and
assume an infinite population of susceptible individuals\footnote{This assumption is largely acceptable 
at the beginning of an epidemic.}. Therefore, we do not need
the \lq correction factor' $1-N(t)/N_0$, and instead use the modulating function
$\mu(t)$ to model the (process-independent) effect of mobility restrictions.
A similar effect due to mobility restrictions can be incorporated into the SIR model, by applying factor $\mu(t)$ to the infection rate $\beta$.   

Given the above connection between a (possibly modulated) Hawkes process
and the traditional SIR, one might ask what is the benefit of our approach
with respect to SIR-like models. 
In contrast to SIR, 
the Hawkes process allows us to choose an arbitrary kernel $\nu(t)$, not necessarily exponential.
With this freedom, can we observe dynamics significantly different from those
that can be obtained by a properly chosen exponential shape? 
We answer this question in the positive with the help of an 
illustrative scenario. 
 
Consider an epidemic starting at time zero with $I_0=1000$
infected individuals.\footnote{Note that in our model the number of immigrants has a Poisson
distribution with mean $\int_0^{\infty} \sigma(t) \diff t$. However to reduce the variance we have preferred 
to start simulations with the same {\em deterministic} number of infected. Further, note that analytical results
for the mean trajectory of infected nodes are not affected by this choice.}   
We fix to $g = 10$ (days) the average generation time, which is 
the mean temporal separation between a new infection belonging to generation $i$ and its parent 
in generation $i-1$.  Note that the above constraint implies
that $\int_0^{\infty} t\,\nu(t) \diff t = g$.

We also normalize $\int_0^{\infty} \nu(t) \diff t = 1$, since we can use $\mu(t)$ to scale the infection rate, in addition to considering time-varying effects due to mobility restriction. Note that in SIR we can only satisfy the above constraints by choosing the exponential kernel $\nu(t) = \frac{1}{g} e^{-\frac{t}{g}}$, $t > 0$. Instead, in our Hawkes model we have much more freedom.
  
\begin{figure}[htb]
\centering
\includegraphics[width=0.9\columnwidth]{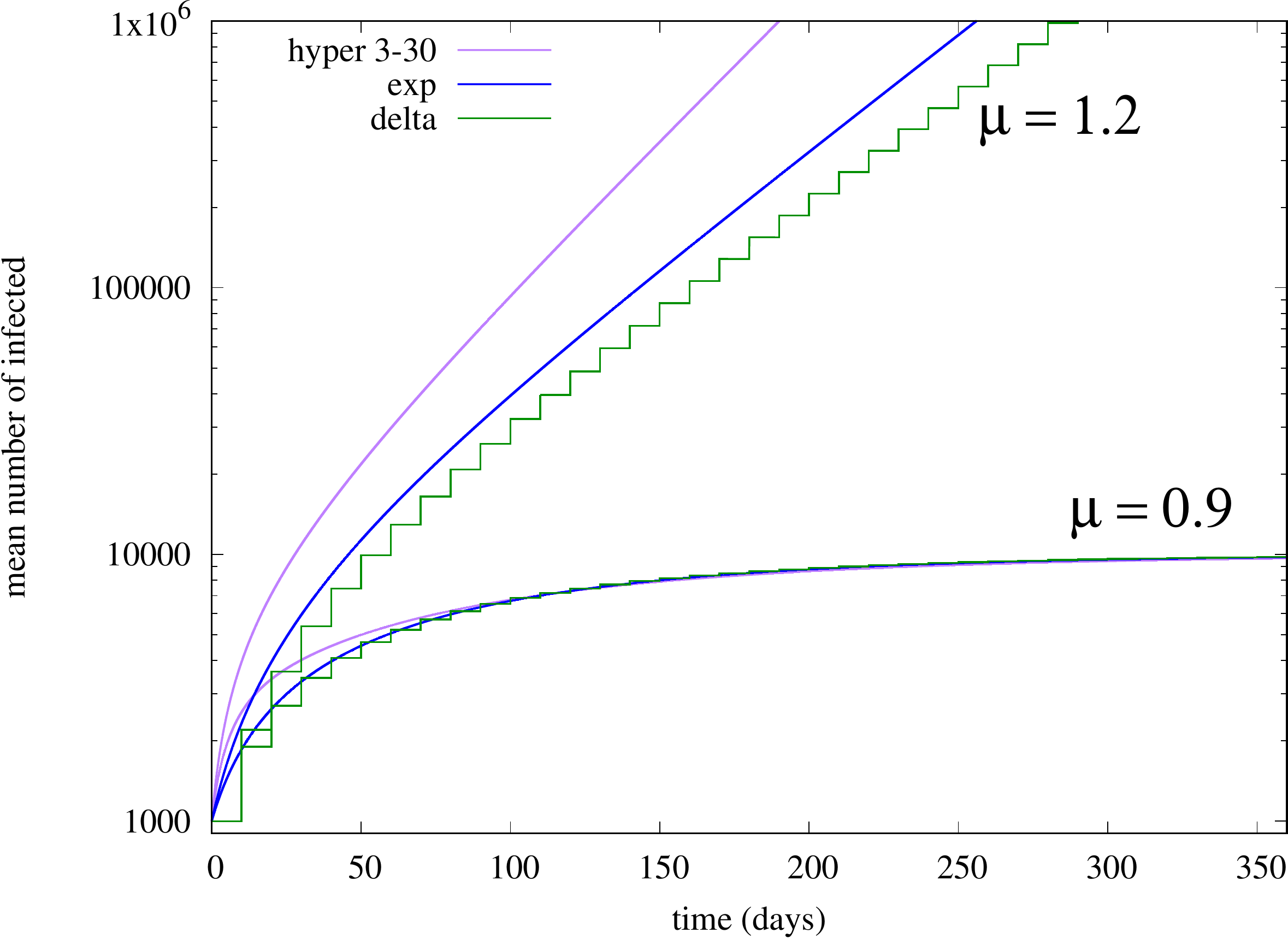}
\caption{Evolution of mean number of infected for different kernel shapes, in the case of constant $\mu = 1.2$ (supercritical)
or constant $\mu = 0.9$ (subcritical), with $I_0 = 1000$, $g = 10$.\label{fig:mucost}}
\end{figure}    
  
First, we consider a scenario in which $\mu(t)$ is constant, equal to either 1.2 (supercritical case) or 0.9 (subcritical case).
In Fig. \ref{fig:mucost} we show the temporal evolution of the mean number of 
infected $\overline{N}(t)$, for three kernel shapes that allows for an explicit solution of \equaref{barlambda},\equaref{barN}
using Laplace transform (\ref{app:explicit}).  
The {\em delta} shape corresponds to the kernel function $\overline{\nu}(t) = \delta(t-g)$, where
$\delta(\cdot)$ is Dirac's delta function, for which
\begin{equation}\label{eq:barNdelta}
\overline{N}_{\rm delta}(t) = I_0 \frac{\mu^{\lfloor t/g \rfloor + 1}-1}{\mu - 1}
\end{equation} 

The {\em exp} shape corresponds to the exponential kernel  $\overline{\nu}(t) = \frac{1}{g} e^{-\frac{t}{g}}$ (SIR model), for which
\begin{equation}\label{eq:barNexp}
\overline{N}_{\rm exp}(t) = I_0 \left[ 1+\frac{\mu}{\mu-1}\left(e^{t (\mu-1)/g} -1\right) \right]
\end{equation} 

The {\em hyper} shape corresponds to the hyper-exponential kernel: \\
\mbox{$\overline{\nu}(t) = p_1 \alpha_1 e^{-\alpha_1 t} + p_2 \alpha_2 e^{-\alpha_2 t}$},
where $0 \leq p_1 \leq 1$, $p_2 = 1 - p_1$, $p_1/\alpha_1 + p_2/\alpha_2 = g$,
which also permits obtaining an explicit, though lengthy expression of $\overline{N}(t)$ that we omit here.
Specifically, the \lq hyper' curve on Fig. \ref{fig:mucost} corresponds
to the case $1/\alpha_1 = 3$, $1/\alpha_2 = 30$.
 

We observe that, in the subcritical case ($\mu = 0.9$), the mean number of infected
saturates to the same value, irrespective of the kernel shape; this can be explained by the fact that
the final size of the epidemics is described by the same branching process for all kernel shapes (i.e., a branching process
in which the offspring distribution is Poisson with mean 0.9).  
In the supercritical case ($\mu = 1.2$), instead, the mean number of infected 
grows exponentially as $\Theta(e^{\eta t})$ (as $t$ grows large), where, interestingly,
$\eta > 1$ depends on the particular shape (notice the log $y$ axes on Fig. \ref{fig:mucost}).
In particular, the hyper-exponential kernel can produce arbitrarily large $\eta$ (\ref{app:explicit}).
We conclude that, even when we fix the average generation time $g$,
different kernels can produce largely different (in order sense) evolutions of $\overline{N}(t)$.
Note that, by introducing compartments, SIR-like models 
can match higher-order moments of the generation time, but our results
suggest that $\overline{N}(t)$ depends on {\em all} moments of it, i.e.,
on the precise shape of the kernel.
Moreover, recall that some shapes are difficult to approximate by a phase-type
approximation (e.g., the rectangular shape, or more in general, kernels with finite support).  
   
\begin{figure}[htb]
\centering
\includegraphics[width=0.9\columnwidth]{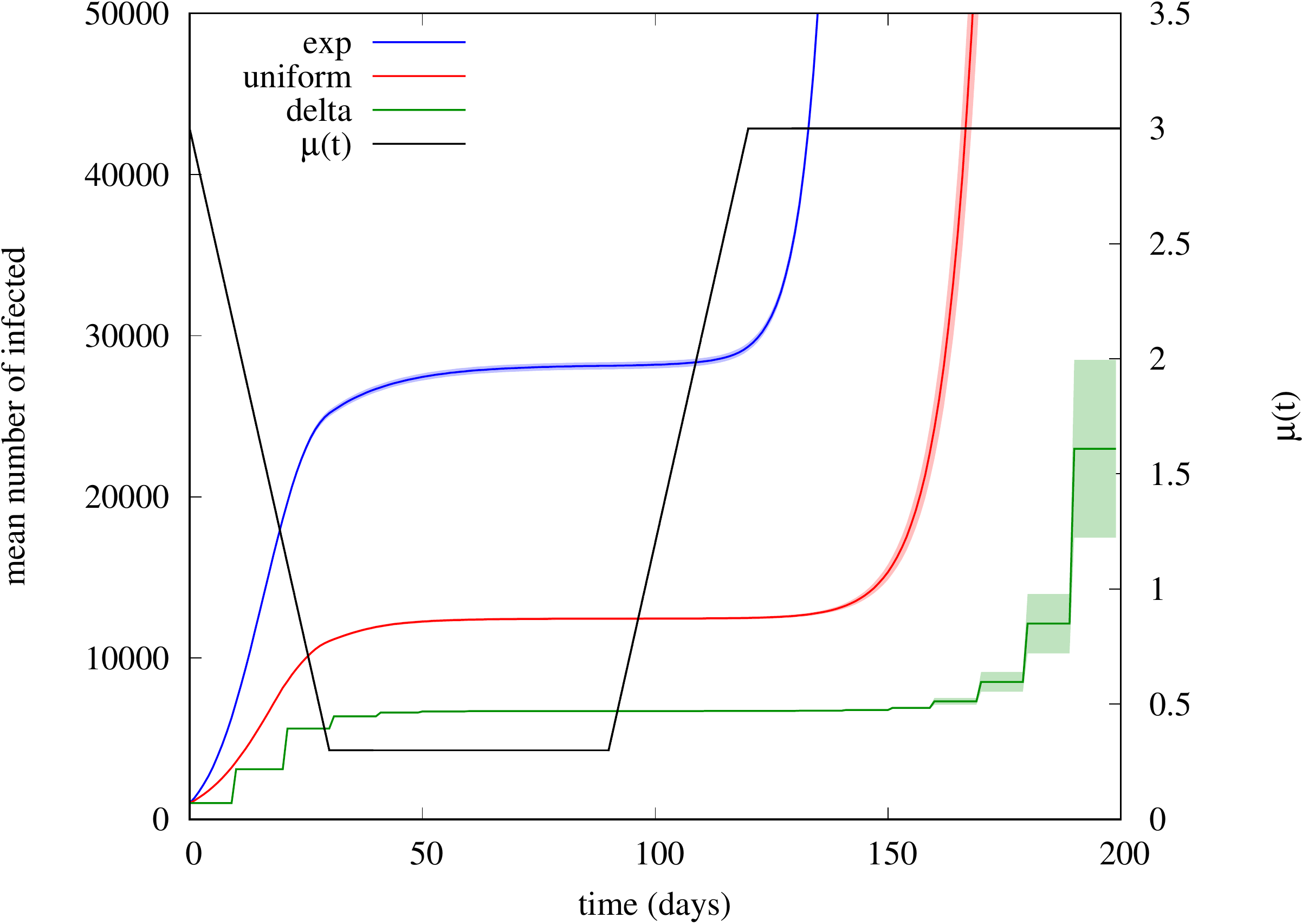}
\caption{Evolution of mean number of infected (left y axes) in the COVID-like scenario, for different kernel shapes. 
Shaded areas denote $95\%-$level confidence intervals obtained by 100 simulation runs. The right y axes 
refers to modulating function $\mu(t)$ (black curve).\label{fig:covcum}}
\end{figure}

The strong impact of the specific kernel shape becomes even more evident
when we consider a time-varying $\mu(t)$, as in our modulated Hawkes process.  
As an example, consider the COVID-inspired scenario in which the modulating function $\mu(t)$ 
corresponds to the black curve in Fig. \ref{fig:covcum}: 
during the first 30 days, $\mu(t)$ decreases linearly from 3 to 0.3; it stays constant at 0.3 for the next 60 days; 
and it goes back linearly to 3 during the next 30 days, after which it stays constant at $3$. 

In Fig. \ref{fig:covcum} we show the mean number of infected estimated
by simulation (averaging 100 independent runs), for the three kernel 
shapes {\em exp}, {\em delta}, and {\em uniform},
where {\em uniform} corresponds to the kernel $\overline{\nu}(t) = \frac{1}{2 g}$, $t \in (0,2g)$,
while the shapes {\em exp}, {\em delta} have been already introduced above.
Shaded areas around each curve denote $95\%-$level confidence intervals.

  




We observe huge discrepancies among the trajectories of $\overline{N}(t)$ obtained
under the three kernels. In particular, after the first \lq wave' of 30 days,
the {\em exp} kernel produces about four times 
more infections than those produced by the {\em delta} kernel. 
This can be explained by the fact that $\mu(t)$ varies significantly on a time window (30 days) 
comparable to the average generation time (10 days).

It is also interesting to compare what happens on the second wave
starting at day 90, after all 3 curves have settled down to an almost
constant value: now discrepancies are even more dramatic: although we observe
a very fast resurgence of the epidemic in all cases, this happens
with significant delays from one curve to another. 
This is due to the fact that the number of individuals who are still infectious
after the subcritical period (when $\mu < 1$) is largely different,
especially considering that both the {\em uniform} and {\em delta} kernels
have finite support (20 and 10 days, respectively) much smaller
than the duration of the subcritical period, whereas the {\em exp}
shape has infinite support: this implies that under the
{\em uniform} and {\em delta} kernels the virus survives the subcritical 
period only through chains of infections belonging to successive
generations, whereas under the {\em exp} shape in principle the
epidemic can restart just thanks to the original immigrants
at time 0, who are still weakly infectious when we re-enter 
the supercritical regime (around day 100).
Actually, in our experiment, under the {\em delta} kernel the epidemic died
out in 68 out of 100 simulation runs, which explains the large confidence
intervals obtained in this case at end of the observation window.
Under the {\em uniform} kernel, the epidemic died only in 3 out of 100 runs,
while it always survived under the {\em exp} kernel.

We conclude that, even while fixing the mean generation time,
the precise shape of the kernel function can play an important role
in predicting the process dynamics and the impact of countermeasures, 
especially when the time-scale of control
is comparable to the time-scale of an individual contagion.
One might obtain a good fitting with measured data also by using an 
exponential shape, and a properly chosen modulating function $\mu(t)$,
but this is undesirable, since the required $\mu(t)$ would no longer
reflect the actual evolution of mobility and interpersonal contact 
restrictions. In the case of COVID-19, several researchers have actually
attempted to incorporate into analytical models detailed 
information about people mobility, using for example data
provided by cellular network operators or smartphone apps \cite{googleapp,appleapp}.        
   
\section{Moment generating function}\label{sec:moments}
Our modulated Hawkes process is stochastic in nature, hence it is important
to characterize how realizations of the process are concentrated around the mean trajectory
derived in Sec. \ref{sec:model}. This characterization is instrumental, for example, 
in designing simulation campaigns with proper number of runs.
Moreover, note that it is entirely possible that the epidemic gets extinct at its early stages,
or in between two successive waves, as we have seen in the scenario in Fig. \ref{fig:covcum}. 
Actually, an epidemic could die out even when starting in the supercritical regime
(consider the case of a single immigrant at time $t$, who does not generate any offspring with probability $e^{-R(t)}$),
something that is not captured by deterministic mean-field approaches. Therefore,
it is interesting to understand the variability of the process at any time.  

Under mild assumptions an expression for the moment generating function of the number of points in $[0,t)$ can be given,
and an iterative procedure can be applied to compute the moment of any order $n$ in terms
of moments of smaller order, though with increasing combinatorial complexity.

In this section we limit ourselves to reporting the main results on this moments' characterization, 
without their mathematical proofs, to keep the paper focused on the application to COVID-19 and 
its control measures.       

Hereon, we shall assume
\begin{equation}\label{hp:boundedh}
K_1:=\sup_{(t,m)\in (0,\infty)\times\mathrm{M}}\nu(t,m)\in (0,\infty),\quad K_2:=\sup_{t\in (0,\infty)}\mu(t)\in (0,\infty).
\end{equation}
Also, we fix a horizon $\tau\in (0,\infty)$, and, for $t\in (0,\tau)$, we denote with $S_{t,\tau}$ the number of points, up to time $\tau$, 
in the cluster generated by an immigrant at $t$ (including the immigrant). 
We denote with 
$|\cdot|$ 
the modulus of a complex number.

We are interested in $N(\tau)$, i.e., the total number of points
generated up to time $\tau$, irrespective of their mark.

\begin{theorem}[Moment generating function of $N(\tau)$] \label{thm:LaplaceTransf}
Assume \eqref{hp:integrability} and 
\eqref{hp:boundedh}. Then there exists $\theta_c>0$ such that, for any $z\in\Theta_c$,
\[
\Theta_c:=\{z\in\mathbb C:\,\,\mathrm{Re}z<\theta_c\},
\]
we have
\begin{equation}\label{eq:LaplaceTransf}
\mathbb{E}[\mathrm{e}^{z N(\tau)}]=\exp\left(\int_{0}^{\tau}(G(t,z)-1)\,\sigma(t)\mathrm{d}t\right)
\end{equation}
and
\begin{equation}\label{eq:LaplaceBound}
\sup_{z\in\Theta_c}|\mathbb{E}[\mathrm{e}^{zN (\tau)}]|<\infty.
\end{equation}
Here
\begin{equation}\label{eq:H}
G(t,z):=\mathbb{E}[\mathrm{e}^{z S_{t,\tau}}],
\end{equation}
is the solution of the following functional equation.
\begin{equation}\label{eq:G}
G(t,z)=\mathrm{e}^{z}\mathbb{E}\left[\mathrm{e}^{\int_{0}^{\tau}(G(v,z)-1)\mu(v)\nu(v-t,M_1)\,\mathrm{d}v}\right],
\quad (t,z)\in (0,\tau)\times\Theta_c.
\end{equation}
\end{theorem} 

From the moment generating function we can obtain an expression for the first and second
moment of $N(\tau)$ (the first moment can also be obtained directly from the average 
stochastic intensity, as we have done in Sec. \ref{sec:model}.).



\begin{theorem}[The first two  moments of $N(\tau)$]\label{thm:law}
Assume \eqref{hp:integrability} and 
\eqref{hp:boundedh}. Then

\begin{align}
\mathbb{E}[N(\tau)]=\overline{N}(\tau)={\int_{0}^{\tau} \overline{ \lambda}(t) \,\mathrm{d}t} =
{\int_{0}^{\tau}E_{1}(t)\sigma(t)\,\mathrm{d}t} \label{eq:1mom}
\end{align}
 and 
\begin{align}
\mathbb{E}[N(\tau)^{2}]
= {\int_{0}^{\tau}E_{2}(t)\sigma(t)\,\mathrm{d}t}+ {\left(\int_{0}^{\tau}E_{1}(t)\sigma(t)\,\mathrm{d}t\right)^2},
\label{eq:2mom}
\end{align}
with
\begin{align}\label{E2}
E_2(t)&=G''(t,0)=\mathbb{E}[S^2_{t,\tau}]= 1+ \int_{t}^{\tau} E_2(v) \mu(v) \overline{\nu}(v-t)\,\mathrm{d}v \,+ \nonumber \\
& + {2}\int_{t}^{\tau}E_{1}(v) \mu(v) \overline{\nu}(v-t)\,\mathrm{d}v + 
 \left(\int_{t}^{\tau} E_{1}(v) \mu(v)\overline{\nu}(v-t) \mathrm{d}v\right)^2
\end{align}


\begin{equation}\label{eq:E1}
E_1(t)=G'(t,0)=\mathbb{E}[S_{t,\tau}]=1+\int_{t}^{\tau}\overline{\lambda^{(t)}}(v)\,\mathrm{d}v
\end{equation}
and
\begin{equation}\label{eq:lambdabar}
\overline{\lambda^{(t)}}(v)=\mu(v)\overline{\nu}(v-t)+
\mu(v)\int_{t}^{v}\overline{\nu}(v-u)\overline{\lambda^{(t)}}(u)\,\mathrm{d}u.
\end{equation}
%
%
\end{theorem}

Note that \eqref{E2} and \eqref{eq:lambdabar} are second-type inhomogeneous Volterra equations.
In the particular case in which $\mu()$ is constant, solutions for \eqref{E2} and \eqref{eq:lambdabar}  
can be found by applying a standard Laplace transform methodology.

\section{COVID-19 Model}\label{sec:covidmodel}
Now we describe how we applied the modulated Hawkes process introduced before
to model the propagation dynamics of COVID-19.  
The proposed model could actually be used to represent the dynamics of other similar 
viruses as well. 

First, we take advantage of the fact that we do not need to 
consider a unique kernel function for all infected.
Indeed, the presence of marks allows us to introduce different classes
of infectious individuals with specific kernel functions.
Specifically, we have considered three classes of infectious: {\em   
symptomatic}, \emph{asymptomatic} and \emph{superspreader}, denoted
by symbols $\{s,a,h\}$. We  assume that, when a person gets infected,
it is assigned a random class  $C \in \{s,a,h\}$ with  probabilities
$p_s,p_a,p_h$, respectively, $p_s + p_a + p_h = 1$, $0 \leq p_s \leq 1$, $0 \leq p_a \leq 1$, $0 \leq p_u \leq 1$.  

As the name suggests, {\em symptomatic} people are those who will develop 
evident symptoms of infection, and we assume that because of that they will be 
effectively quarantined at home or hospitalized at the onset of symptoms. 
On the contrary, the \emph{asymptomatic} mark is given to individuals 
who will not develop strong enough symptoms to be quarantined. 
Therefore, they will be able to infect other people for the entire duration of the disease,
though at low infection rate (unless they get scrutinized by mass testing, as explained later). 
At last, \emph{superspreaders} are individuals who exert a high infection rate but do not get quarantined
due to several possible reasons (unless they get scrutinized by mass testing). 
This class also includes people with mild symptoms, who 
become highly contagious because of their mobility pattern (e.g., participation to \lq superspreading events').
Though the above classification of infectious individuals is a simplified one,
a properly chosen mix of the three considered classes can represent a wide range 
of different scenarios.   

Irrespective of their class, we will assume that all infectious people 
go through the following sequence of stages: first, there is
a random incubation time, denoted by r.v. $I$, with given cdf $F_I()$. 
During this time we assume the all infected exert a low infection rate $\lambda_{\rm low}$. 
Then, there is a crucial pre-symptomatic period, during which
infected in classes $\{s,h\}$ already exert a high infection rate $\lambda_{\rm high}$,
while infected in class $a$ still exert low infection rate $\lambda_{\rm low}$. 
For simplicity, we have assumed the presymptomatic phase to have constant duration $w$.  

The following evolution of the infection rate of an individual depends on the class:
since we assume that {\em symptomatic} people get effectively quarantined, 
they no longer infect other individuals after the onset of symptoms. We model this fact 
introducing a {\em quarantined} class of people (denoted by $q$),
and deterministically moving all infected in class $s$ to class $q$ after time $I+w$. 
  
People in classes $\{a,h\}$ continue to be infectious during a disease period of random
duration $D$, with given cdf $F_D()$. The difference between these two classes is that infectious 
in class $a$ ($h$) exert, during the disease period, infection rate $\lambda_{\rm low}$ ($\lambda_{\rm high}$), respectively.
At last, we assume that people in classes $\{a,h\}$ enter a residual period
of random duration $E$, with cdf $F_E()$, during which all of them exert infection rate $\lambda_{\rm low}$.
We introduced this additional phase because some people
who recovered from COVID-19 where found to be still contagious several days (even weeks) after
the end of the disease period. 
Note that durations $I,D,E$ are assumed to be independent, and that 
the complete mark associated to an infected is the tuple $M=(C,I,D,E)$.   

\begin{figure}[htb]
\centering
\includegraphics[width=0.9\columnwidth]{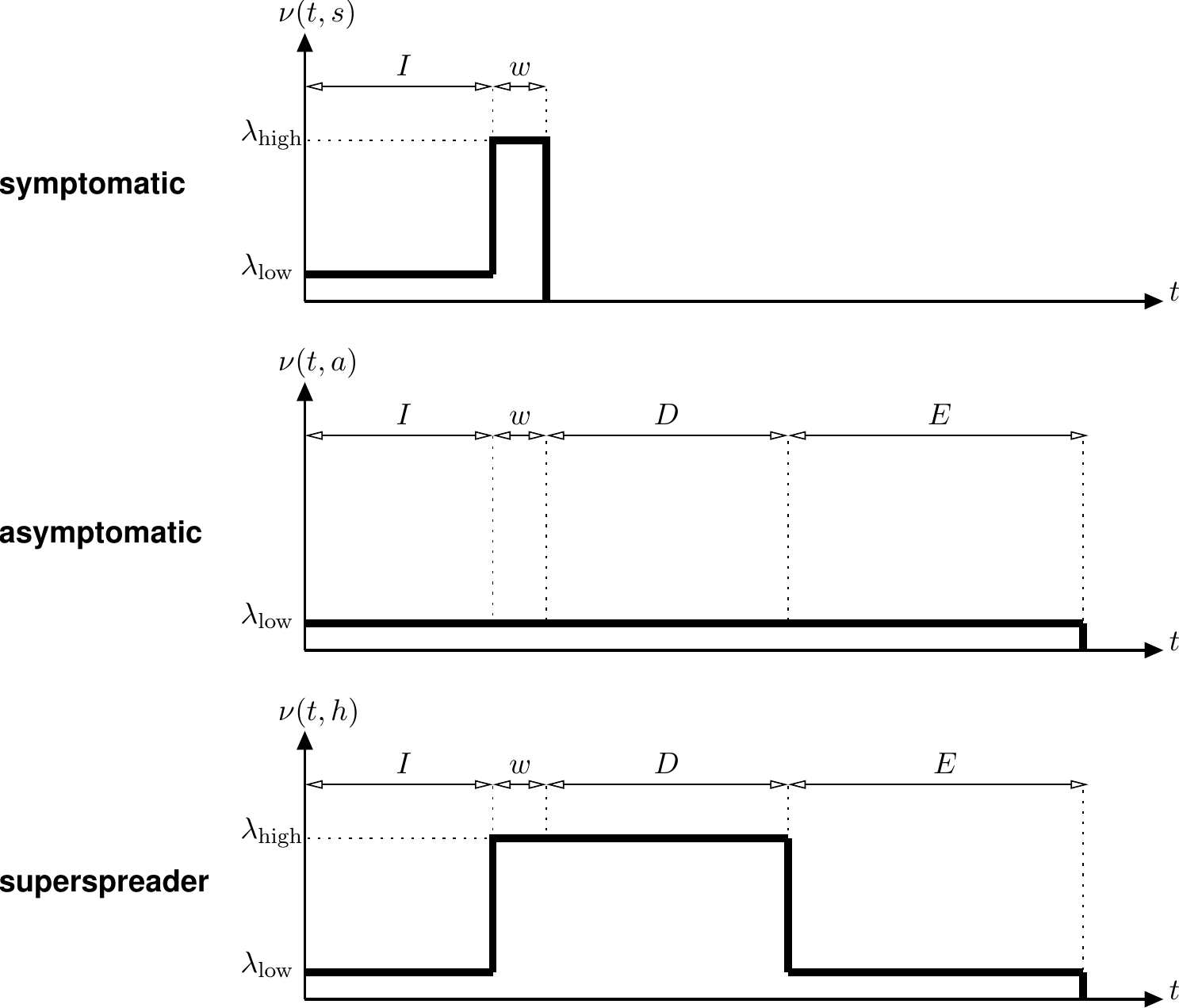}
\caption{Kernel functions for classes $s,a,h$ (from top to bottom), for given durations $I,D,E$ of stages. \label{fig:fert}}
\end{figure} 

An illustration of the three class-dependent kernels $\nu(t,s),\nu(t,a),\nu(t,h)$, conditioned
on the durations $(I,D,E)$, is reported in Fig. \ref{fig:fert}. 

\begin{remark}
We emphasize that our model of COVID-19 is targeted at predicting the process of new infections,
rather than the current number of people hosting the virus in various conditions. In particular, we do not explicitly
describe the dynamics of symptomatic but quarantined people and their exit process 
(i.e., recovery or, in the worst case, death), since such dynamics have no effect on the
spreading process (under the assumption of perfect quarantine). 
However, if desired, one could describe through appropriate probabilities and time distributions
how quarantined people split between those isolated at home and those who get hospitalized,
the fraction going to intensive care, and those who unfortunately die.    
\end{remark}

\begin{table}[tbhp]
\centering
\begin{tabular}{|c|c|c|} 
 \hline
 Parameter & symbol & COVID-19 fitted value \\ 
 \hline \hline
 incubation period  & I & tri([2,12], mean 6) \\
 \hline
 pre-symptoms period & w & 2 \\
 \hline
 disease period & D & unif([2,12])\\
 \hline
 residual period & E & exp(10) \\
 \hline
 low infection rate & $\lambda_{\rm low}$ & 0.05 \\
 \hline
 high infection rate &  $\lambda_{\rm high}$ & 1\\
 \hline
 symptomatic probability & $p_s$ & 0.06 \\
 \hline
 asymptomatic probability & $p_a$ & 0.91\\
 \hline
 super-spreader probability & $p_h$ & 0.03\\
 \hline 
\end{tabular}
\caption{Virus-specific parameters.\label{tab1}}
\end{table}

The parameters introduced so far, summarized in Table \ref{tab1}, are related 
to specific characteristics of the virus. 
We now model properties of the specific environment
where the virus spreads, taking into account the impact of countermeasures.
First, we need to specify the immigration process $\sigma(t)$.
To keep the model as simple as possible, we have assumed that
the system starts at a given time with $I_0$ new infections, i.e.,
immigrants arrive as a single burst concentrated at one specific instant.


The impact of countermeasures is taken into account in the model in two
different ways. First, modulating function $\mu(t)$ can be used to model 
the instantaneous reduction of the infection rate at time $t$ due to the current
mobility restrictions, and, more in general, changes in the environment
which affect the ability of the virus to propagate in the susceptible population
(such as seasonal effects). Typically, $\mu(t)$ is a decreasing 
function in the initial part of the epidemics, in response to 
new regulations introduced by the government, while it goes up again
when mobility restrictions are progressively released. 

Second, we assume that any infected not already found to be positive
is tested (e.g., by means of massive swab campaigns) at individual, instantaneous rate $\rho(t)$,
which reflects both the amount of resources employed by the health service
to discover the infected population, and the effectiveness of tracking.
Recall that, when an infected is found positive, we assume it transits to the
quarantined state $q$ and stops infecting other people.
 
\begin{figure}[htb]
\centering
\includegraphics[width=0.9\columnwidth]{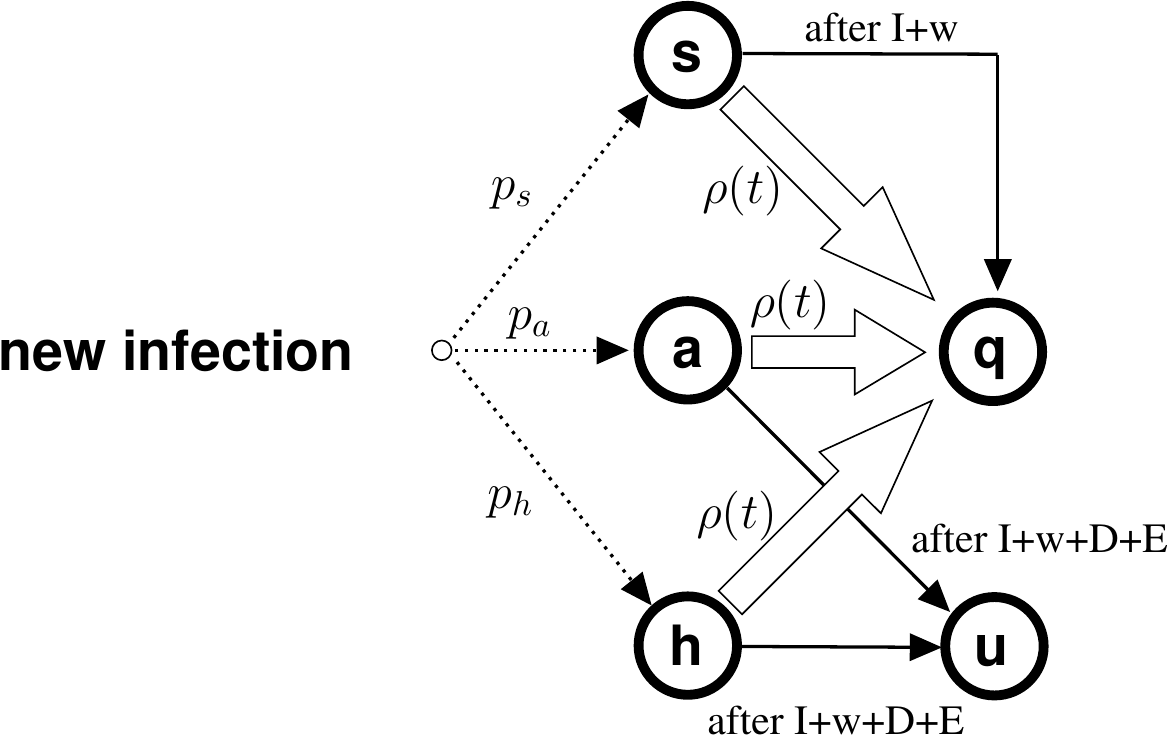}
\caption{State transitions of an infected individual.\label{fig:mark}}
\end{figure}
 
Fig. \ref{fig:mark} shows the resulting transitions that can occur for the 
different classes. Note that class $q$ collects all cases
known to the health authorities at a given time $t$, and thus coincides with the 
set of detected cases. Infected in classes $s,a,h$ are still unknown
to the health service (but they are detectable). When they stop to be positive,
people in classes $a,h$ transit to a class $u$ (undetected), which collects
all infected who remain unknown to the health service.   

Environment-related parameters are summarized in Table \ref{tab2}.

\begin{table}[tbhp]
\centering
\begin{tabular}{|c|c|c|} 
 \hline
 Parameter & symbol & fitted value for first wave in Italy \\ 
 \hline \hline
 initial day & $-\Delta t$ & -20 \\ 
 \hline
 initial number of infected & $I_0$ & 370 \\ 
 \hline
 modulating function & $\mu(t)$ & $T_a = -3$, $\mu_a = 3.71$ \hspace{19mm} \\
 & & $T_b = 30$, $\mu_b = 0.31$,   (see Fig. \ref{fig:muitaly}) \\
 \hline
 detection rate & $\rho(t)$ & $\rho(t) = 0.000115 \,t$, \quad (see Fig. \ref{fig:muitaly}) \\ 
 \hline 
\end{tabular}
\vspace{5mm}
\caption{Environment-specific parameters.\label{tab2}}
\end{table}

\begin{remark}
We emphasize that the model described so far does not explicitly model spatial effects and 
sub-populations. As such, it is more suitable to describe a homogeneous 
scenario, where its environmental parameters (including 
modulating functions $\mu(t)$ and detection rate $\rho(t)$) can be reasonably assumed
to apply to all individuals of the population irrespective of their spatial
location. The natural candidate of such scenario is a single nation
or just a sub-region of it. By so doing, we can assume that common national
regulations are applied, as well as common health-care practices.
However, our approach based on Hawkes processes could be extended to incorporate
spatial effects, by considering spatio-temporal kernel functions $\nu({\bm x},t)$, where
${\bm x}$ is a spatial vector originated at the point at which a new infection occurs.
{\color{black} Another possibility would be to consider a multivariate
temporal Hawkes process, where each component represents a homogeneous region,
and the different regional processes can interact with each other.} 
\end{remark}

\subsection{Computation of the real-time reproduction number $R(t)$}
From our model, it is possible to compute in a native way the average 
number of infections caused by an individual who gets infected at time $t$, i.e.,  
the real-time reproduction number $R(t)$. 
To compute $R(t)$, we condition on the duration $x$ of the incubation time,
on the duration $y$ of the disease time, on the duration $z$ of the residual time, and on the class assigned to the node
getting infected at time $t$ (note that they are all independent 
of each other):
\begin{multline}\label{eq:Rt2}
R(t) = \int_x \int_y \int_z \Bigg( 
\lambda_{\rm low} \int_{0}^{x} \mu(t+\tau) u(t,\tau) \diff \tau + 
p_a \lambda_{\rm low} \int_{x}^{x+w+y+z} \mu(t+\tau) u(t,\tau) \diff \tau + \\
(p_s+p_h) \lambda_{\rm high} \int_{x}^{x+w} \mu(t+\tau) u(t,\tau) \diff \tau + 
p_h \lambda_{\rm high} \int_{x+w}^{x+w+y} \mu(t+\tau) u(t,\tau) \diff \tau + \\
p_h \lambda_{\rm low} \int_{x+w+y}^{x+w+y+z} \mu(t+\tau) u(t,\tau) \diff \tau \Bigg)  \diff F_E(z) \diff F_D(y) \diff F_I(x)
\end{multline}
where
$$ u(t,\tau) = e^{-\int_t^{t+\tau} \rho(s) \diff s} $$
is the probability that a node which gets infected at time $t$ is still undetected
at time $t+\tau$.

\section{Model fitting for the first wave of COVID-19 in Italy}\label{sec:firstwave}
We fit the model to real data related to the spread of COVID-19 in 
Italy, publicly available on GitHub \cite{ministero}.
Italy was the country where the epidemics first
spread outside of China into Europe, causing about 34600 deaths
at the end of June 2020 during the first wave. 

Our main goal was to match the evolution of the number of detected cases,
represented in the model by individuals in class $q$.
The actual count is provided by the Italian government on a daily basis 
since February 24th 2020. We take this date as our reference day zero.
However, it is largely believed that the epidemics started well before the 
end of February. 
Indeed, it became soon clear that detected cases were just the top of a much bigger iceberg, as the prevalence of asymptomatic infection was initially 
largely unknown, which significantly complicated the first modeling 
efforts to forecast the epidemic evolution. 
During June-July 2020, a blood-test campaign (aimed at detecting IgG antibodies) was conducted on a representative population of 64660 people to understand the actual diffusion of the first wave of COVID-19 in Italy \cite{serologico}.
As a main result, it has been estimated that 1 482 000 people have been 
infected. 
This figure provides a fundamental hint to properly
fit our model, and indeed while exploring the parameter space
we decided to impose that the total number of predicted cases at the end
of June (day 120) is roughly 1 500 000. 
   
For the durations $I$,$w$,$D$,$E$ of the different stages of an infection,
we have tried to follow estimates in the medical literature.
In particular, the duration of the incubation period is believed to 
range between 2 and 12 days, with a sort of bell shape 
around about 5 days \cite{incubation,jcm9020538,Qineabc1202}, 
which we have approximated, for simplicity,
by a (asymmetric) triangular distribution with support $[2,12]$ and mean 6.
Moreover, we have fixed the duration of the pre-symptomatic phase to 2 days.
For the duration of the disease period we have taken a uniform
distribution on $[2,12]$, while the residual time is modeled by an 
exponential distribution with mean 10 days \cite{Byrne2020}. 

The other virus-related parameters have been set as reported in 
the third column of Table \ref{tab1}. Note that, since we
further apply the external modulating function $\mu(t)$, we could
arbitrarily normalize to 1 the value of $\lambda_{\rm high}$, while we set $\lambda_{\rm low} = 0.05$.

For what concerns environment-related parameters (see Table \ref{tab2}),
a first problem was to choose a proper initial day $-\Delta t$ (recall
that day zero is the first day of the trace) at which to start the process
with $I_0$ initially infected individuals.
While different pairs $(\Delta t, I_0)$ are essentially equivalent,
we decided not to start the process with too few cases and too much 
in advance with respect to day 0, to limit the variance
of a single simulation run. We ended up setting $\Delta t = 20$, while 
$I_0 = 370$ was selected as explained later.

\begin{figure}[htb]
\centering
\includegraphics[width=0.9\columnwidth]{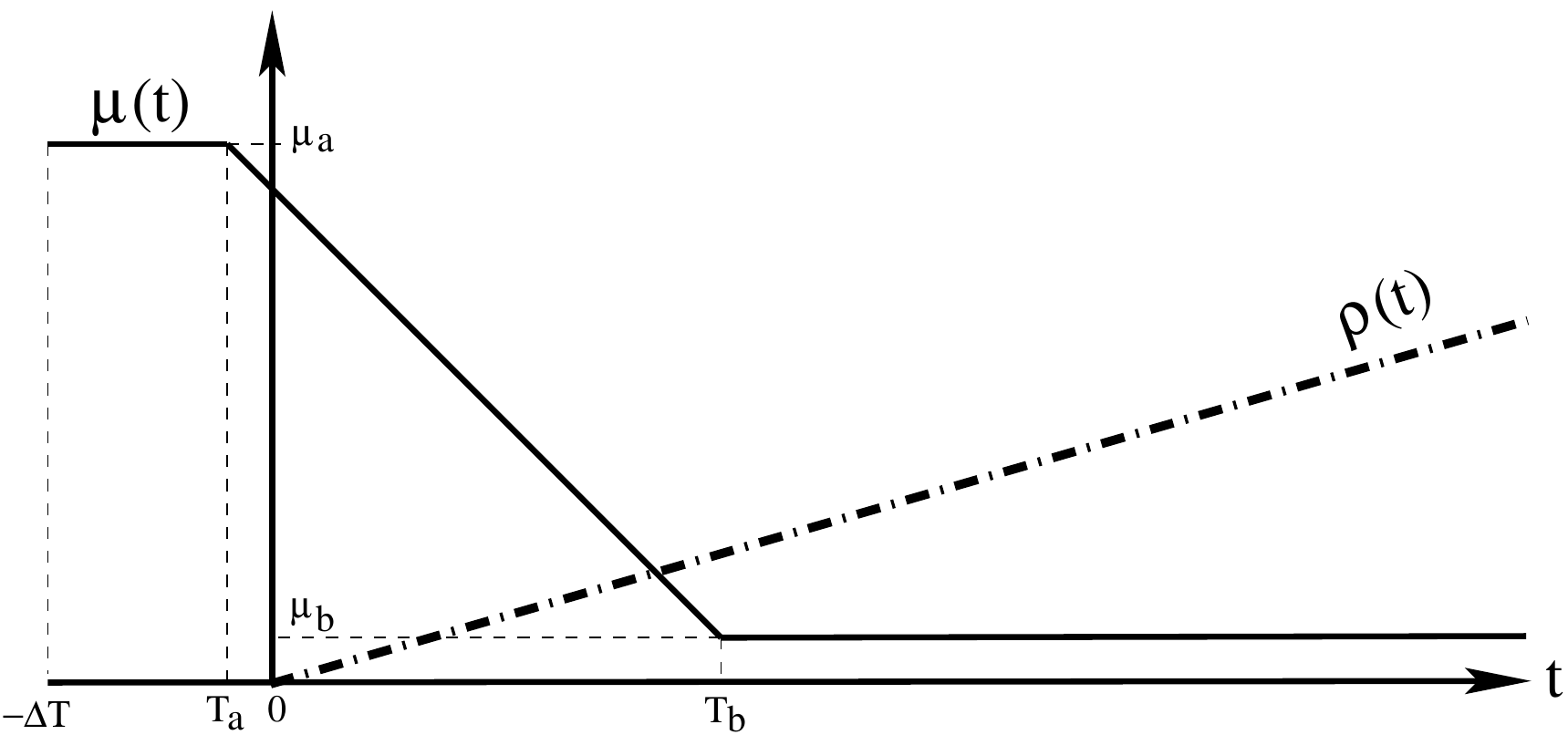}
\caption{Chosen profiles and parameters of functions $\mu(t)$ and $\rho(t)$ for the first wave of COVID-19 in Italy.\label{fig:muitaly}}
\end{figure}

Mobility restrictions in Italy were progressively enforced by national laws
starting a few days before day 0, first limited to red zones in Lombardy, and soon extended 
to the entire country through a series of increasingly restrictive regulations, 
introduced over the next 30 days. Instead of trying the capture the step-wise 
nature of such restrictions, we assume $\mu(t)$ to take the simpler 
profile depicted in Figure \ref{fig:muitaly}, i.e., a high initial value
$\mu_a$ before time $T_a$, a low final value $\mu_b$ after time $T_b$,
and a linear segment connecting point $(T_a,\mu_a)$ to point
$(T_b,\mu_b)$. We decided to set $T_a = -3$ and $T_b = 30$ to reflect the
time window in which mobility restriction were introduced.


In Fig. \ref{fig:muitaly} we also show our choice for the profile of detection rate $\rho(t)$, i.e.,
a linear increase starting from day 0, with coefficient $\alpha$,
$\rho(t) = \max\{0,\alpha t\}$. This profile is justified by the 
fact that the first wave caught Italy totally unprepared ($\rho = 0$ before day zero), 
while massive swabs were only gradually deployed over time after day zero. 
  
The critical parameters $I_0,\mu_a,\mu_b,\alpha$ were fitted by a 
minimum mean square error (MMSE) estimation technique based on the
curve of detected cases in the time window $[0-120]$ days, 
after all other parameters were manually selected as detailed above.

\myfig{10}{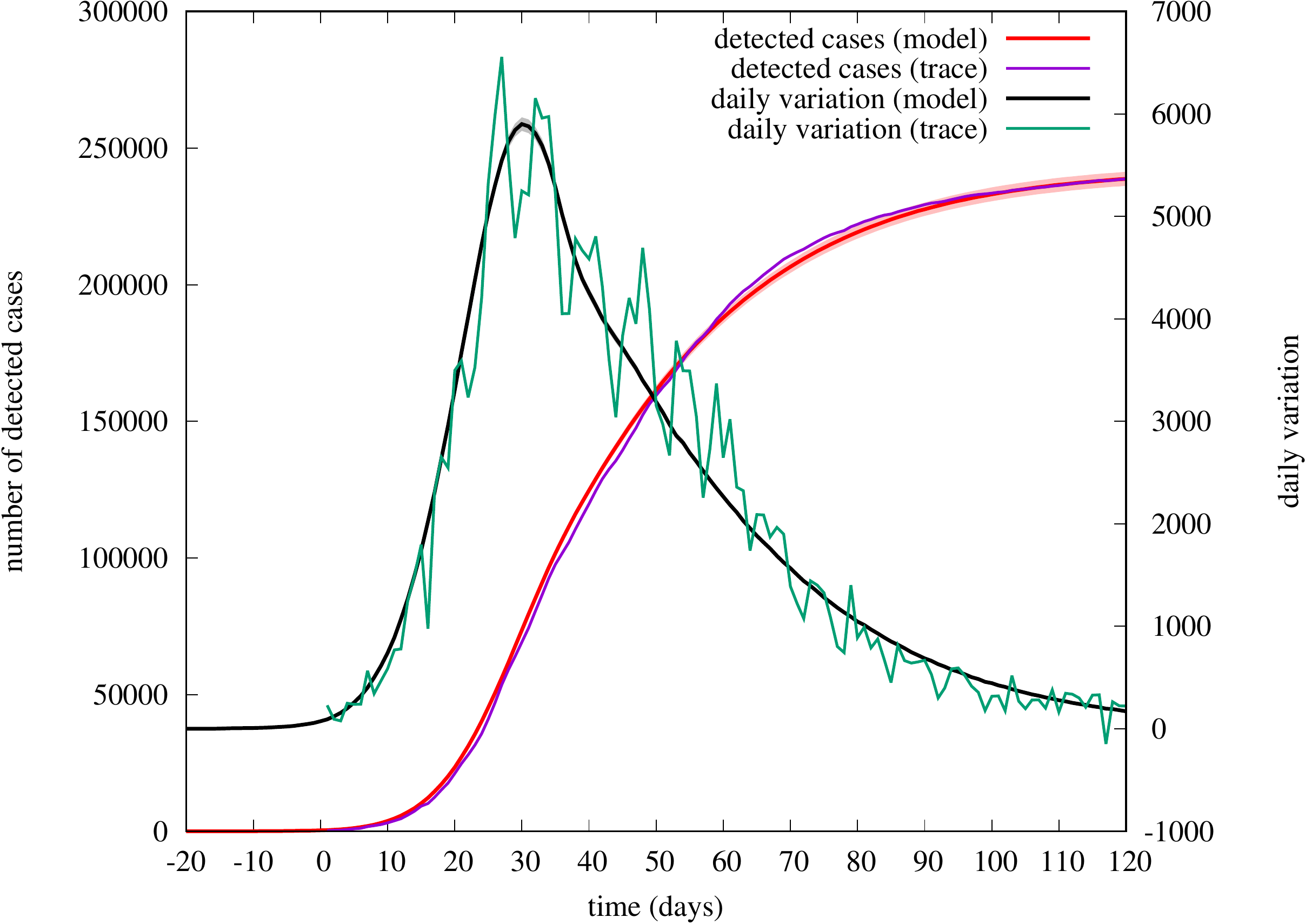}{Evolution of the number of detected cases according to model and real data. Cumulative number (left y axes)
and its daily variation (right y axes).}

\myfig{10}{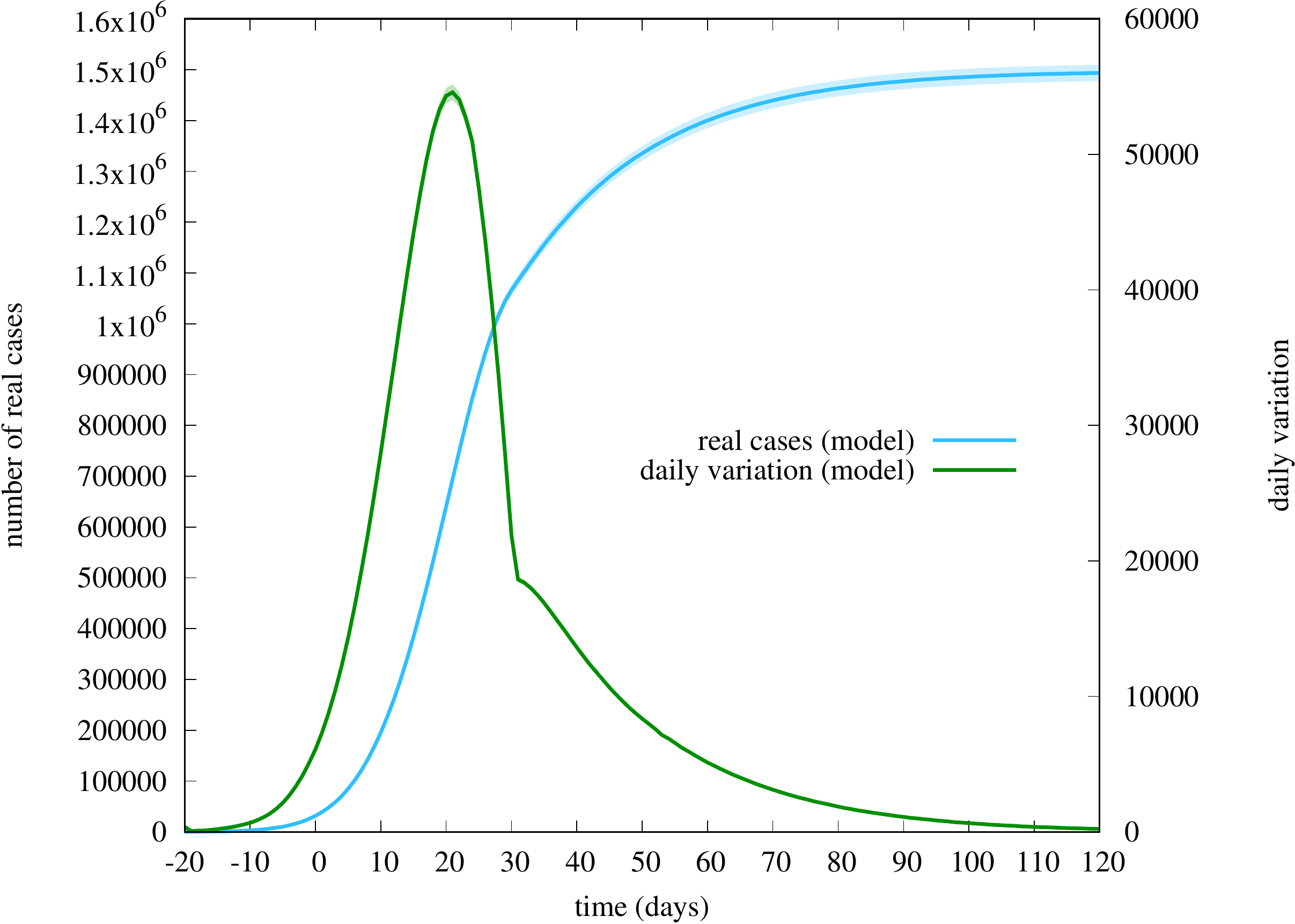}{Evolution of the actual number of cases according to model only. Cumulative number (left y axes)
and its daily variation (right y axes).}

Fig. \ref{fig:deteconf} shows the final outcome of our fitting,
comparing the evolution of the number of detected cases according to model 
and real data. We show both the cumulative number (left $y$ axes) and its daily increment (right $y$ axes). 
Analytical predictions for the mean trajectory of detected cases 
were obtained by averaging 100 simulation runs. The shaded region around
analytical curves (barely visible) shows 95\%-level confidence intervals
computed for each of the 120 days. 

We emphasize that a similar good match could be possibly obtained by other sets of 
parameters. Our purpose here was not to compute the best possible fit
in the entire parameter space (which would be nearly impossible), but to show
that the model is rich enough to capture the behavior observed on the real trace
after a reasonable choice of most of its parameters, driven by their physical meaning.

Fig. \ref{fig:totconf} shows instead the evolution of the real number of 
cases (both the cumulative number and its daily variation) according to the model
only, in the absence of data. Note however that we have constrained ourselves to 
obtain a total number of about 1 500 000 cases on day 120, as suggested by the
serological test \cite{serologico}.

Interestingly, by looking at the values of $\mu_a$ and $\mu_b$ computed by our MMSE, 
it appears that the national lockdown was able to reduce the spreading 
ability of the virus within the Italian population by a factor of 
about 12 (from 3.71 to 0.31). We will see later on in Fig. \ref{fig:reverse} 
that our fitting of $\mu(t)$ for the first wave is consistent with 
mobility trends estimated from the usage of the Apple maps application \cite{appleapp}.

\myfig{10}{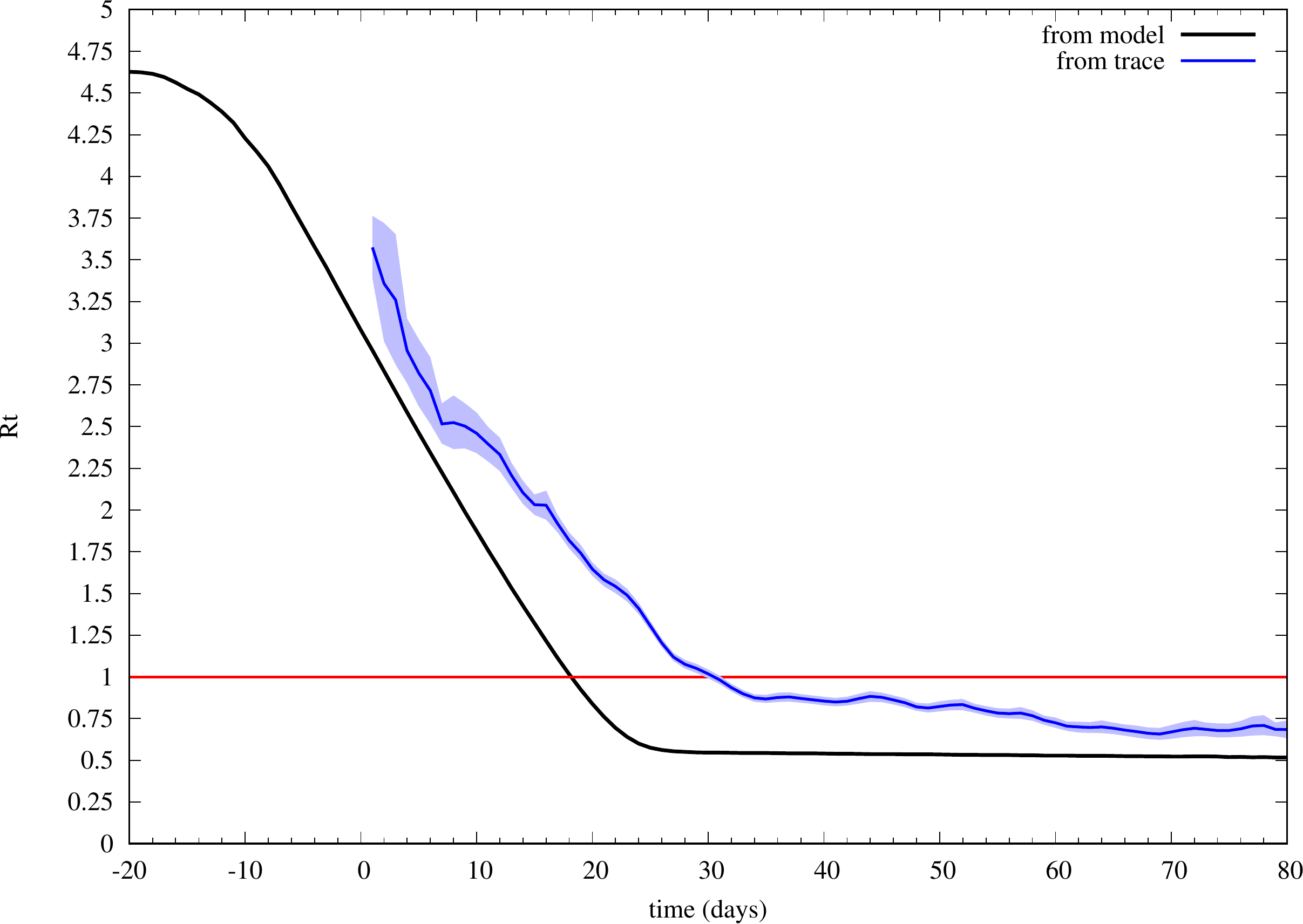}{Real-time reproduction number computed by the model, compared to its 
estimation based on the trace of detected case, according to the Wallinga-Teunis method 
(with 95\% confidence interval denoted as shared region).}

In Fig. \ref{fig:Rtconf} we show the real-time reproduction number computed by \equaref{Rt2}.
We also applied the classic method of Wallinga-Teunis \cite{wallinga}, implemented as in the R0 package \cite{Obadia2012},
to estimate $R(t)$ from the trace of detected cases. To apply their method, we used a generation time
obtained from a Gamma distribution with mean 6.6 (shape 1.87, scale 0.28), which has been proposed 
for the first wave of COVID-19 in Italy \cite{cereda2020early}.
Interestingly, though both estimates of $R(t)$ exhibit qualitatively the same behavior,
the model-based value of $R(t)$, which considers also undetected cases, tends to be smaller.

\subsection{What-if scenarios}
Having fitted the model to the available trace, we proceed to exploit the model to examine interesting what-if
scenarios.
First, we investigate what would have happened (according to the fitted model)
if lockdown restrictions were shifted in time by an amount of days $\delta$.
This means that we keep all parameters the same, except that we translate horizontally in time the profile
of $\mu(t)$ depicted in Fig. \ref{fig:muitaly}.
In Figure \ref{fig:whatif1} we show the total number of cases that would have 
occurred for values of $\delta \in \{-7,-3,0,3,7\}$.

\myfig{10}{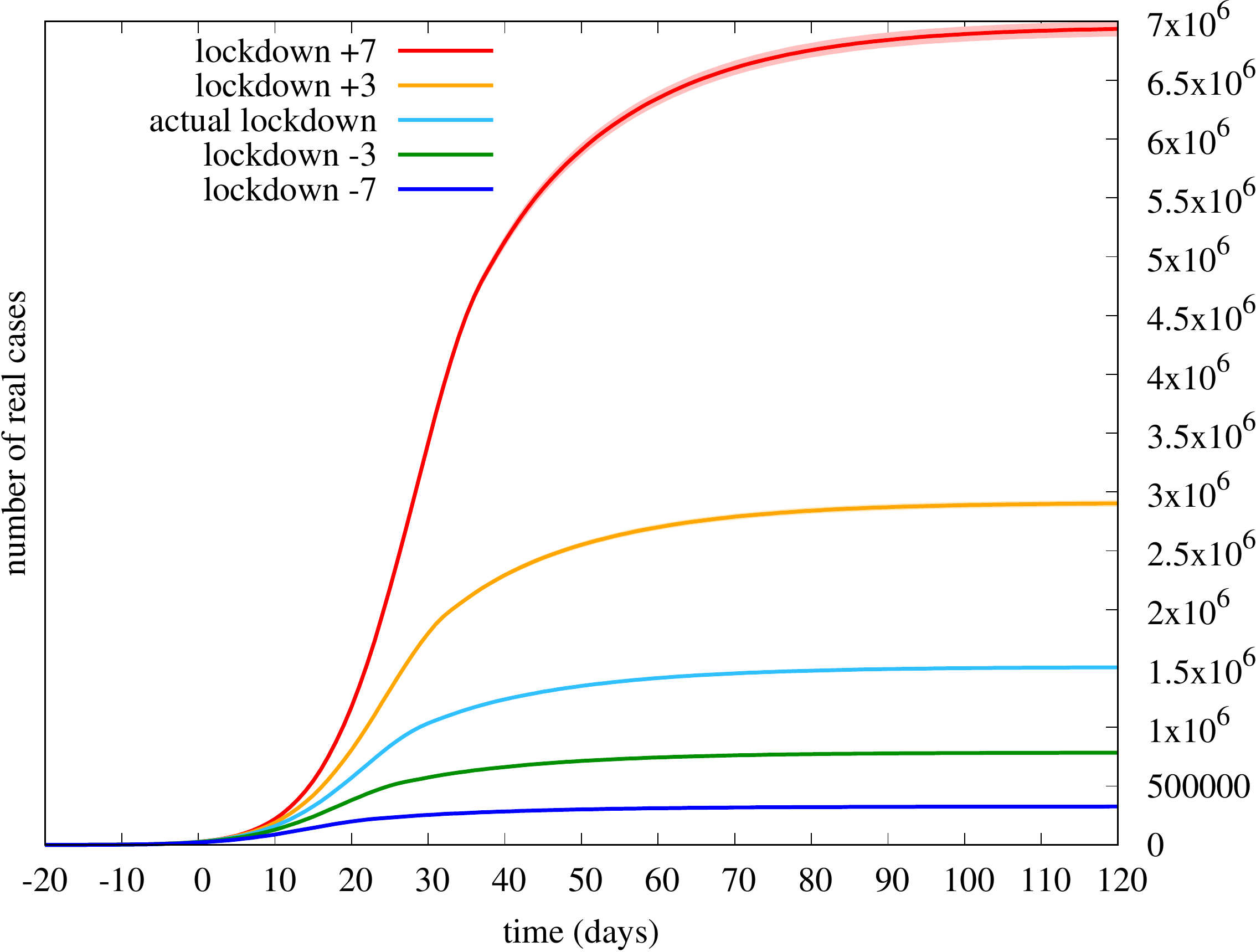}{What-if scenario: total number of cases if restrictions were shifted in time by $\delta$ days.}

We observe that a shift of just 3 days corresponds to a factor of roughly 2
in the number of cases. This translates, dramatically, into an equivalent impact
on the number of deaths, if we assume that the mortality rate would have stayed the same\footnote{This
is somehow optimistic, since mortality also depends on the saturation level of intensive therapy facilities.} 
(i.e., $34600/1500000 \sim 2.3\%$). In other words, a postponement (anticipation) of lockdown restrictions
by just 3 days wold have caused twice (half) the number of deaths, which is a rather impressive result.
   
\myfig{10}{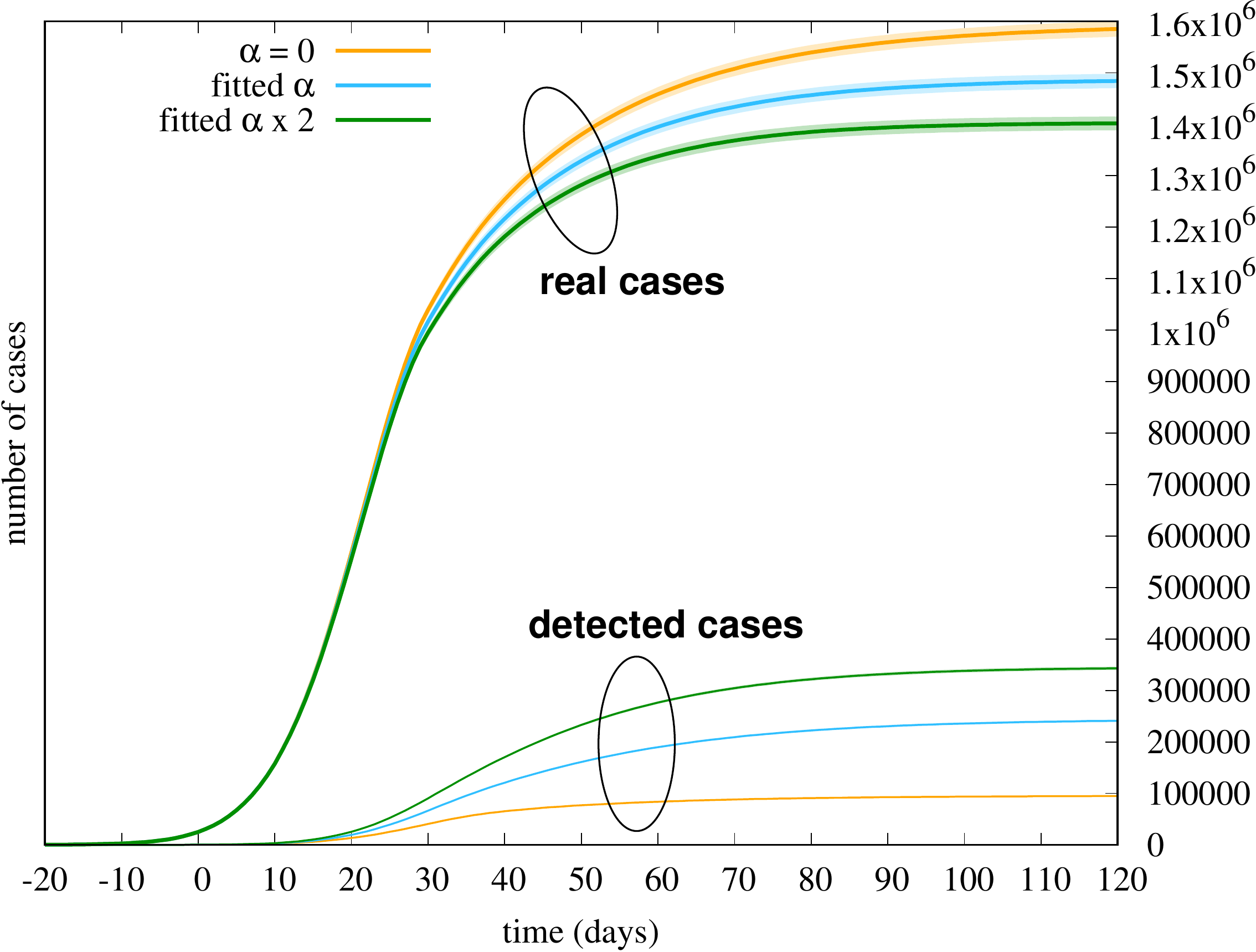}{What-if scenario: impact of detection rate $\rho(t) = \alpha t$ on the
real and detected number of cases, varying $\alpha$.}   
   
In Figure \ref{fig:whatif2} we investigate instead the impact of detection rate $\rho(t)$, by changing its slope $\alpha$ (recall
Fig. \ref{fig:muitaly}). We report both the number of real cases and the number of detected cases predicted by the 
model, when all other fitted parameters are kept the same. We consider what would have occurred with $\alpha = 0$ (which 
means that infectious people are never tested), and by doubling the intensity of the detection rate
(a tracking system twice more efficient). 
As expected, with $\alpha = 0$ only symptomatic cases ($p_s = 6\%$) are eventually
detected. 
This time the effect on the final number of cases (or deaths) is not as dramatic as in the previous what-if scenario. 
This suggests that the impact of mass testing in Italy during the first wave  
was marginal, and doubling the efforts would not have produced significant changes in the final 
outcome.

\section{Model fitting for the second wave of COVID-19 in Italy}\label{sec:secondwave}

The second wave of COVID-19 hit Italy in late summer 2020, as in many other European 
countries, mainly as en effect of relaxed mobility in July/August and possibly 
other seasonal effects.
It is interesting to compare the second wave with the first wave, by looking at the 
daily increments of detected cases and deaths, see Fig. \ref{fig:backfromdeath}.

\myfig{12}{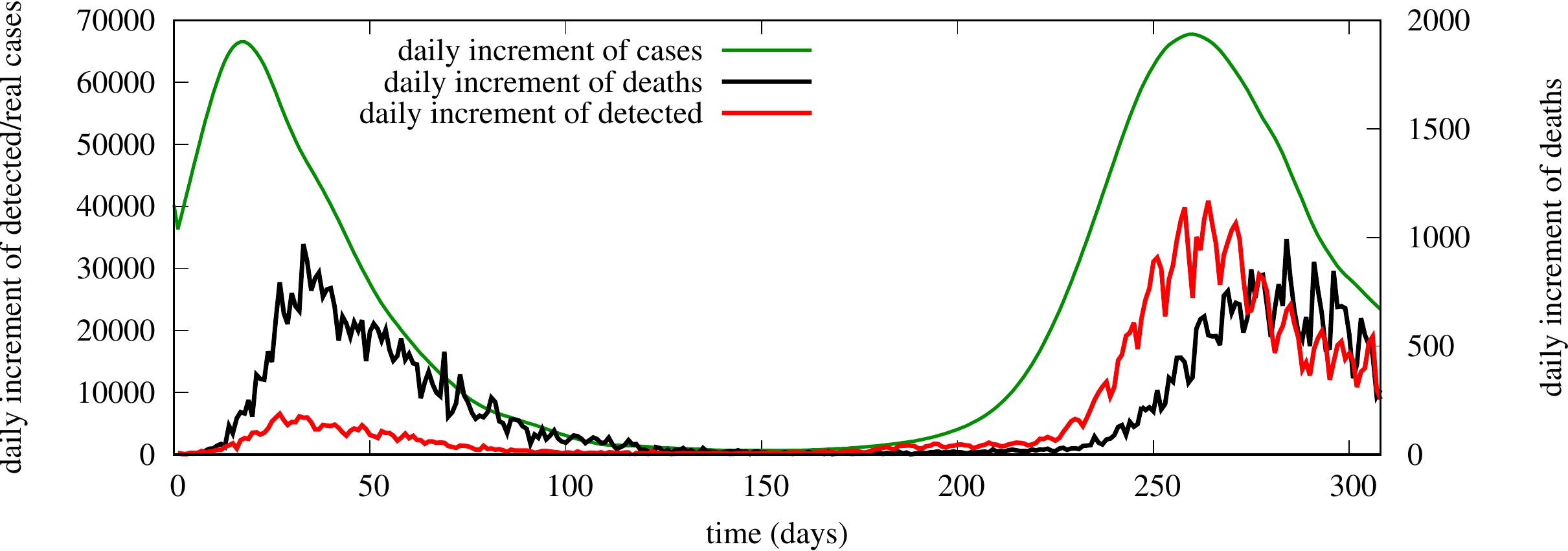}{Daily increment of the number of deaths (black curve, right $y$-axes),
daily increment of the number of detected (red curve, left $y$ axes),  
and estimated daily increment of real cases 
projected back from the curve of deaths (green curve, left $y$-axes).}  

We observe that the number of daily deaths is similar between the two waves. The daily increment of detected cases, instead, is 
very different (around 5000 at the peak of the first wave, around 35000 at the peak of the second wave, a 7-fold increase). This can be explained by the much larger capacity of the health service to perform swabs and track down the infected, built on the experience gained
from the first wave. It also suggests that, differently from the first wave, the impact
of $\rho(t)$ (the individual rate at which an infectious is detected) is expected to be much 
more important during the second wave than in the first wave, 
requiring a careful treatment of it in the model.

We have indeed tried to fit our model to the second wave in Italy,   
keeping all parameters related to the virus (previously fitted for the first wave)
unchanged (Table \ref{tab1}), and adapting only environment-specific 
parameters (Table \ref{tab2}). A major difficulty that we had to face was the unknown (at the time of writing)
actual diffusion of the virus in the population during the second wave. Recall that, 
for the first wave, we exploited a blood-test campaign to get a reference 
for the total number of real cases at the end of the first wave.
For the second wave we employed instead a different approach based on a projection
back in the past of the increment of deaths. A similar idea has been adopted
in \cite{Flaxman2020} to estimate the time-varying reproduction number
in different European country at the onset of the first wave.  
  
Specifically, we got from \cite{deathstat}
the indication of the median (11 days) and IQR (6-18 days) of the amount of time
from symptoms onset to death during Oct-Dec 2020, that we fitted by a Gamma distribution
(shape 1.65, scale 8.45). By convolving such Gamma distribution with the distribution of incubation time, and the pre-symptoms period, we obtained a distribution of the total time
from infection to death, that we used to estimate the time in the past at which 
each dead person was initially infected. By amplifying the number of
infections leading to death by the inverse of the mortality rate 
we eventually obtained an estimate of the daily increment of real cases, as
reported in Fig. \ref{fig:backfromdeath} (green curve, left $y$-axes).

We chose August 1st (day 160 on Fig. \ref{fig:backfromdeath}) as starting
date of the new infection process producing the second wave in Italy.   
This time, instead of manually searching for suitable profiles of $\mu(t)$ and $\rho(t)$
generating the expected curves of real and detected cases, we adopted 
a novel \lq\lq reverse-engineering" approach: we took the daily increments of real
and detected cases as input to the model, and computed the functions $\mu(t)$ and $\rho(t)$
that would exactly produce in the model the given numbers of real and detected cases, on each day\footnote{Discretized
values of functions $\mu(t)$ and $\rho(t)$ are uniquely determined in the model, once we constrain ourselves to  
produce a given number of real and detected cases at each time instant.}.  
  
\myfig{10}{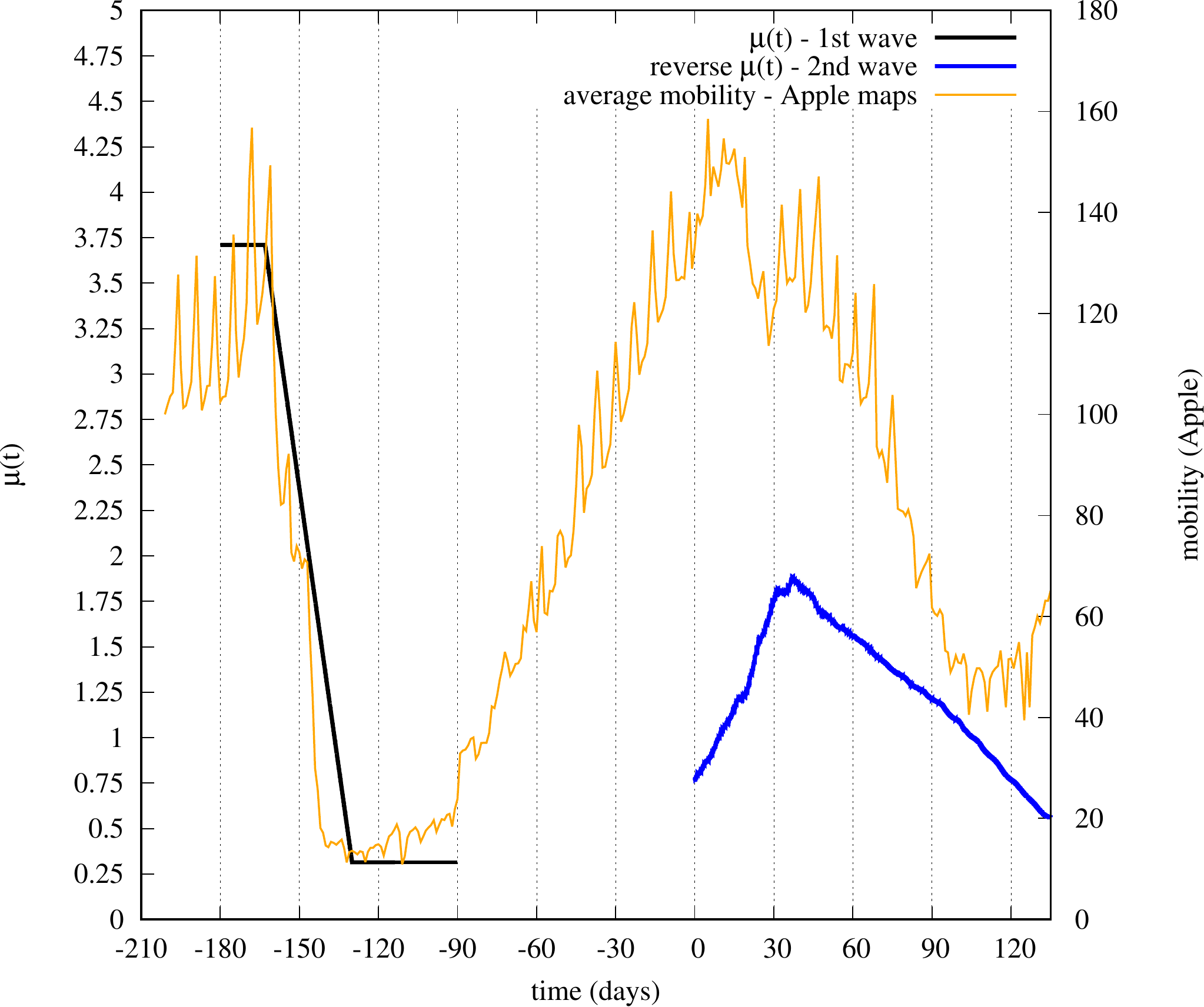}{Modulating functions $\mu(t)$ for the first wave (black, fitted) and second wave (blue, reversed), 
and average mobility according to Apple maps application.}  

In Fig. \ref{fig:reverse} we show the obtained \lq reversed' $\mu(t)$ (blue curve), starting 
from day 0 (August 1st) of the time reference adopted for the second wave.
We also report the average mobility measured in Italy by the Apple maps application \cite{appleapp},
where we have given equal weight (1/3) to driving, transit and walking mobility.
We observe that the \lq reversed' $\mu(t)$ qualitatively follows the same behavior
of mobility measured on the maps application, with a gradual increase during the month of August
followed by a gradual decrease as people (and the government) started to react to the 
incipient second wave.     

For completeness, we have also reported on Fig. \ref{fig:reverse} the fitted $\mu(t)$ for the first wave (black curve).
We observe that, during the first wave, the quick introduction of hard lockdown
caused an abrupt decay of both measured mobility and fitted $\mu(t)$, characterized
by a bigger reduction in a shorter time. The second wave, instead, was characterized by a
smoother transition, due to the different choice of applying just partial lockdowns
and progressive restrictions more diluted over time\footnote{The fact that, during the second wave,
mobility values similar to those of the first wave do not translate into equally similar
values of $\mu(t)$ can be attributed to increased awareness of people during the second wave.}.      

In Fig. \ref{fig:rho} we show instead the \lq reversed' function $\rho(t)$, focusing on the
second wave. We also report on the same plot the daily increments
of real and detected cases, which allow us to better understand the obtained profile of $\rho(t)$,
highlighting an interesting phenomenon occurred around day 30.
Indeed, we observe that, during the first month, the epidemic was closely tracked
by the national health system, but at some point, around day 30, the curve of detected cases 
stops increasing, and stays roughly constant during the entire second month, lagging more and more behind the
otherwise exploding curve of real cases (time window 30-60). As consequence, function $\rho(t)$, which
describes the effectiveness of individual tracking, falls down to a minimum reached at around day 60.
This behavior can be interpreted as an effect of the saturation of the
capacity to perform swabs, resulting in a progressive collapse of the tracking system,
as actually experienced by many people during those days.
         
\myfig{10}{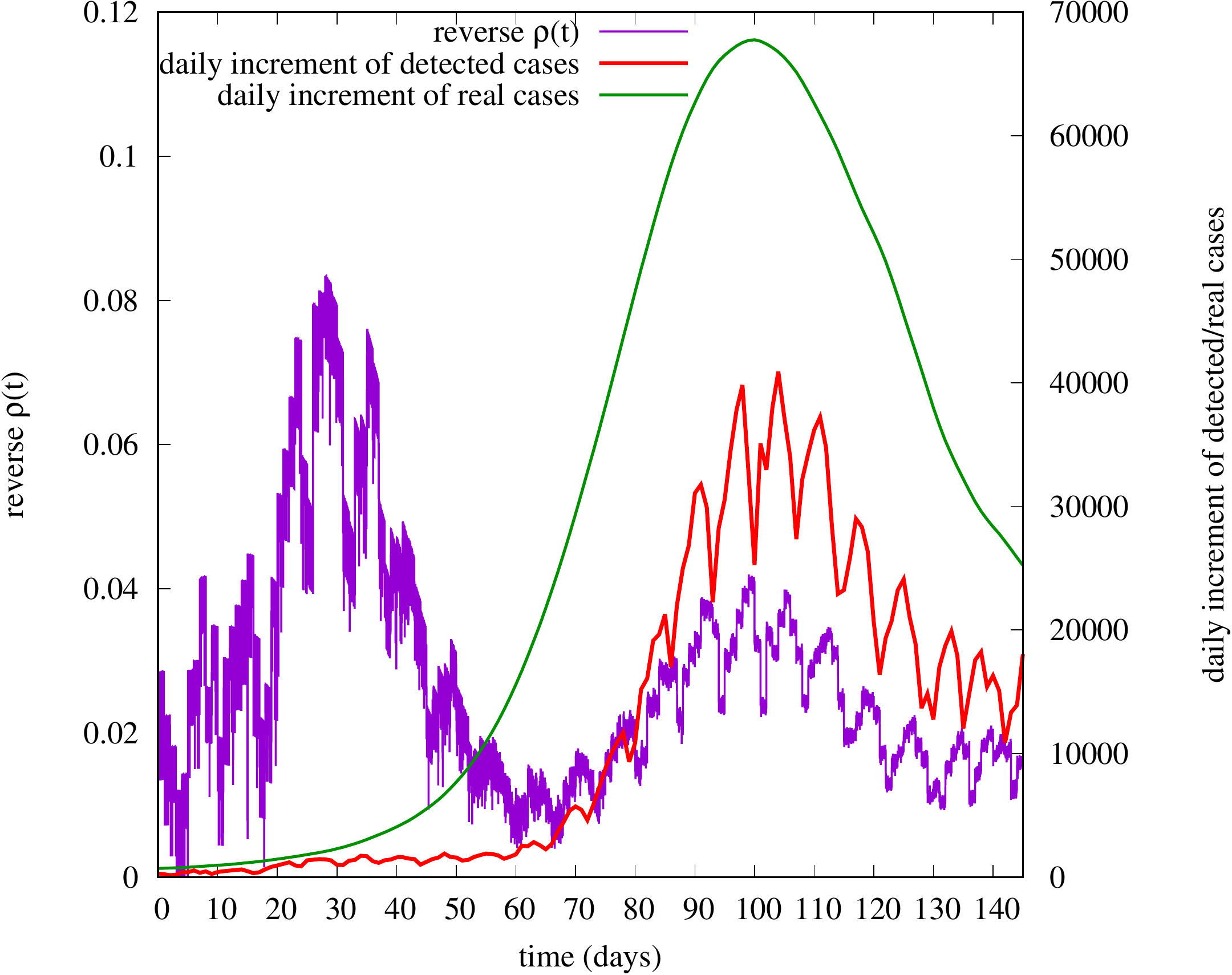}{Reverse detection rate $\rho(t)$ for the second wave (left y-axes), 
and daily increase of real and detected cases (right y-axes).}  

Our analysis shows that, in contrast to what might be believed by just looking at the curve of detected cases, 
September (days 30-60) was, perhaps, the most critical period for the outbreak
of the second wave of COVID-19 in Italy. During this period, the detection capacity of the
national health system was saturated, and could not keep the pace
with the rapid growth of real cases, giving instead the illusion of maintaining the epidemic
under control. 
Our reverse-engineering approach can thus shed some light on what actually happened
at the onset of the second wave, and {\em quantitatively} assess the collapse
of the tracking system.

\section{Conclusions}\label{sec:conclusions}

We have proposed a time-modulated version of the Hawkes process
to describe the temporal evolution of an epidemic within an infinite population 
of susceptible individuals. Our approach allows us to take into account precise distributions 
for the time spent in different stages of the infection, which is of paramount importance 
when the intensity of countermeasures (mobility restrictions, testing and tracing)
varies significantly on time-scales comparable to that of an individual infection.
We have applied the model to the spread of COVID-19 in Italy, 
either by a direct fit of its parameter (first wave), or by a novel reverse 
fit (second wave) which allows us, in retrospect, to understand from data the time-varying 
effectiveness of applied countermeasures. 
Future work will extend the model to overcome some of its
current limitations, like the impact of spatial effects
and other sources of heterogeneity in the population, such as age groups.
We think the proposed approach is promising and could be usefully applied  
to explain the epidemic progress and forecast/assess the impact
of control/mitigation measures.

\appendix

\section{Explicit solution of $\overline{\lambda}(t)$ for constant $\mu(t) = \mu$}\label{app:explicit}
When $\mu(\cdot)\equiv\mu$ is constant, denoting with  $*$  the convolution product,
we can rewrite equation \equaref{barlambda} as
\[
\overline{\lambda}(t)=\sigma(t)+\mu\cdot(\overline{\nu} *\overline{\lambda})(t),\quad t>0,
\]
which can be easily solved in the transformed domain. 
For a non-negative function $f:\mathbb{R}_+\to\mathbb{R}_+$, we denote by
\[
\widehat{f}(s):=\int_{\mathbb{R}_+}\mathrm{e}^{-st}f(t)\,\mathrm{d}t,\quad s\in\mathbb{C}
\]
the Laplace transform of $f$. 
Then
we have:
\[
\widehat{\overline{\lambda}}(s)= \frac{\widehat{\sigma}(s)}{1- \mu\widehat{\overline{\nu}}(s)} \qquad \text{for } \mathrm{Re}( s) >  \mathrm{Re}  (z_{\max}),
\]
with  $z_{\max}$ equal to the  zero of  $1- \mu\widehat{\overline{\nu}}(s)$ with largest real part.

In addition, formally, as long as $\mu <1$ and $\sigma$ is bounded, we can write
\[
\overline{\lambda}(t)= \sigma(t)+\sigma(t)*\sum_{i=1}^{\infty} \mu^i \overline{\nu}^{*i}(t)
\]
where $\overline{\nu}^{*i}(t) $ is the $i$-th fold convolution of $ \overline{v}(t)$.

An analytical expression of $\overline{\lambda}(t)$ can be obtained when $\sigma(s)$ and $\overline{\nu}(s)$ are both  rational.
Table \ref{tab:barlambda} reports the dominant\footnote{\color{black} We recall that  a non negative function $f(t)$  is said to be $\Theta(1)$, iff there exist two constants $0<c \le C <\infty$ such that $c\le  f(t) \le C$, for sufficiently large $t$.} term of $\overline{\lambda}(t)$ (and also of $\overline{N}(t)$) for the case in which $\sigma(t)=\beta\exp(-\beta t)$, and $\overline {\nu}(t)$ takes different shapes satisfying: 
$$ \int_0^{\infty} \overline{\nu}(t) \diff t = 1 \quad;\quad \int_0^{\infty} t\,\overline{\nu}(t) \diff t = \frac{1}{\alpha} $$

\begin{table}
\hspace{-10mm}
\begin{tabular}{|c|c|c|}
\hline
  shape & $\overline{\nu}(t)$ & $\overline{\lambda}(t)$\\
\hline 
{\em delta} & $\delta(t-\frac{1}{\alpha})$ & $\Theta(1)\,  {\mathrm e}^{\alpha \log(\mu) t}$ \\
{\em erl-}$2$ &  $(2\alpha)^2 t{\mathrm e}^{-2\alpha t } $ & $\Theta(1)\, {\mathrm e}^{2\alpha( \sqrt{\mu}   -1) t }$  \\
{\em exp} & $\alpha {\mathrm e}^{-\alpha t } $ & $ \Theta(1)\, {\mathrm e}^{ \alpha(\mu-1)t} $ \\
{\em hyper}$_1$ &    $\frac{{2z^2}{\mathrm e}^{-\frac{2z}{z+1}\alpha t }+2{\mathrm e}^{-\frac{2}{z+1}\alpha t }}{(z+1)^2}$
   & $\Theta(1)\, {\mathrm e}^{\alpha t \left[ \frac{(\mu -1)(1+z^2) -2 z + \sqrt{( \mu^2 (1+z^2)^2  + (z^2-1)^2 (1-2\mu)}}{(1+z)^2} \right]}$ \\
  {\em hyper}$_2$ & $ \frac{z^3\alpha {\mathrm e}^{- z \alpha t }+  \alpha  e^{- \frac{\alpha}{z} t } }{z(z+1)}$
 & $\Theta(1)\,{\mathrm e}^{\alpha t\frac{(\mu-1)(1+z^2) - \mu z +  \sqrt{\mu^2z^2+( z - 1)^2[\mu^2(z^2+1) -2\mu(z^2+z+1) + (z+1)^2 ] } }{2z} }$\\ 
\hline
\end{tabular}
\caption{Dominant term of $\overline{\lambda}(t)$ for different kernel shapes having the same 
average generation time $g = 1/\alpha$.  \label{tab:barlambda}}
\end{table}

Besides the simple deterministic ({\em delta}) and exponential ({\em exp}) shapes, we consider
the Erlang-2 ({\em erl-}$2$) and two variants of hyper-exponential, whose general form is 
$\overline{\nu}(t) = p_1 \alpha_1 e^{-\alpha_1 t} + p_2 \alpha_2 e^{-\alpha_2 t}$. 
To reduce the degrees of freedom of the general hyper-exponential, 
we have assumed a particular relationship between $p_1/p_2$ and $\alpha_1/\alpha_2$, which allows us
to introduce a single parameter $z$. 
Specifically, we have set either 
$p_1/p_2=\alpha_1/\alpha_2=z$  (denoted by {\em hyper}$_1$) or $p_1/p_2=\sqrt{\alpha_1/\alpha_2}=z$ (denoted by {\em hyper}$_2$). 
Note that in both cases the variance of the corresponding hyper-exponential distribution increases with $z \ge 1$,
becoming arbitrarily large as $z\to \infty$.   

For all kernel shapes, and $\mu > 1$, the growth rate of $\overline{\lambda}(t)$ is exponential, i.e., $\overline{\lambda}(t) = \Theta(e^{\eta t})$,
and the same holds for $\overline{N}(t) = \int_0^t  \overline{\lambda}(\tau) \diff \tau$.
The rate $\eta$ of the exponential growth, however, depends on the specific shape.

In particular, for the two considered cases of hyper-exponential shape, $\eta$ is an increasing function of $z$:
while in the {\em hyper}$_1$ case $\eta$ saturates to $2(\mu-1)\alpha$, as $z \rightarrow \infty$,
in the {\em hyper}$_2$ case $\eta$ is even unbounded, as $z \rightarrow \infty$.

We also notice that the rate of the exponential kernel is larger than the rate of the Erlang-2 kernel,
which is in turn larger than the rate of the deterministic kernel.

These results confirm that, even under a constant modulating function $\mu(t)$, the epidemic growth rate
depends on the specific shape of the kernel function $\overline{\nu}(t)$.

%
%
%
%



\bibliography{mybibfile}

\end{document}